# Temporal Symmetry in Primary Auditory Cortex: Implications for Cortical Connectivity

Jonathan Z. Simon,[1,2,3,4] Didier A. Depireux,[5] David J. Klein,[1,3] Jonathan B. Fritz,[3] and Shihab A. Shamma[1,3,4]

[1]*Department of Electrical and Computer Engineering*

[2]*Department of Biology*

[3]*Institute for Systems Research*

[4]*Program in Neuroscience and Cognitive Sciences*

  *University of Maryland*

  *College Park MD 20742–3311, USA*

[5]*Anatomy and Neurobiology*

  *University of Maryland*

  *Baltimore MD 21201 USA*

*Running head*: Temporal Symmetry and Cortical Connectivity

*Contact*: Jonathan Z. Simon, Department of Biology, University of Maryland,

      College Park MD 20742–3311; jzsimon@eng.umd.edu; 301-405-3645

## Abstract

Neurons in primary auditory cortex (AI) in the ferret (*Mustela putorius*) that are well described by their spectro-temporal response field (STRF), are found also to have a distinctive property that we call temporal symmetry. For temporally symmetric neurons, every temporal cross-section of the STRF (impulse response) is given by the same function of time, except for a scaling and a "Hilbert rotation". This property held in 85% of neurons (123 out of 145) recorded from awake animals, and in 96% of neurons (70 out of 73) recorded from anesthetized animals. This property of temporal symmetry is highly constraining for possible models of functional neural connectivity within and into AI. We find that the simplest models of functional thalamic input, from the ventral Medial Geniculate Body (MGB), into the entry layers of AI are ruled out because they are incompatible with the constraints of the observed temporal symmetry. This is also the case for the simplest models of functional intracortical connectivity. Plausible models that do generate temporal symmetry, both from thalamic and intracortical inputs, are presented. In particular, we propose that two specific characteristics of the thalamo-cortical interface may be responsible. The first is a temporal mismatch between the fast dynamics of the thalamus and the slow responses of the cortex. The second is that all thalamic inputs into a cortical module (or a cluster of cells) must be restricted to one point of entry (or one cell in the cluster). This latter property implies a lack of correlated horizontal interactions across cortical modules during the STRF measurements. The implications of these insights in the auditory system, and comparisons with similar properties in the visual system, are explored.

# *INTRODUCTION*

The Spectro-temporal Response Field (STRF) is one measurement used to describe how an auditory neuron responds to a spectrally dynamic sound (Aertsen & Johannesma, 1981b; Eggermont, Aertsen, Hermes, & Johannesma, 1981; Hermes, Aertsen, Johannesma, & Eggermont, 1981; Johannesma & Eggermont, 1983; Smolders, Aertsen, & Johannesma, 1979). It is related to the spectral "response area" of a neuron, defined roughly as the range of frequencies and intensities of pure tones that elicit excitatory or inhibitory responses (though response area measurements are blind to timing information within the responses). The STRF is a measure of the spectral and dynamic properties of auditory response areas, and has been used in a variety of auditory areas in both mammals and birds, and using a variety of stimuli ranging from simple, e.g. auditory ripples, to complex, e.g. natural sounds (Aertsen & Johannesma, 1981a; deCharms, Blake, & Merzenich, 1998; Epping & Eggermont, 1985; Escabi & Schreiner, 2002; Fritz, Shamma, Elhilali, & Klein, 2003; Kowalski, Versnel, & Shamma, 1995; Linden, Liu, Sahani, Schreiner, & Merzenich, 2003; Miller, Escabi, Read, & Schreiner, 2002; Qin, Chimoto, Sakai, & Sato, 2004; Rutkowski, Shackleton, Schnupp, Wallace, & Palmer, 2002; Schafer, Rubsamen, Dorrscheidt, & Knipschild, 1992; Schreiner & Calhoun, 1994; Sen, Theunissen, & Doupe, 2001; Shamma & Versnel, 1995; Shamma, Versnel, & Kowalski, 1995; Theunissen, Sen, & Doupe, 2000; Valentine & Eggermont, 2004; Yeshurun, Wollberg, & Dyn, 1987).

[Figure 1 about here.]

The stimuli and techniques, adapted from studies of visual processing (De Valois & De Valois, 1988), and from psychoacoustic research (Green, 1986; Hillier, 1991; Summers & Leek, 1994), apply linear and nonlinear systems theory to measure the response area of auditory units. From the systems theoretic point of view, the spike train output (averaged over many trials) is determined by the spectro-temporal profile of a broadband, dynamic, acoustic input (Depireux, Simon, & Shamma, 1998). As can be seen from the example in Figure 1, the STRF includes quantitative information shaping how the neuron determines its firing rate, as a function of both spectrum (a vertical cross-section gives firing rate as a function of frequency at some moment in time), and time (a horizontal cross-section gives firing rate as a function of time, for a single frequency).

There are also analogs of the STRF in other sensory modalities. The most prominent are in vision, where the Spatio-Temporal Response Field (De Valois & De Valois, 1988) partially inspired the study of auditory STRFs, and the somatosensory domain (DiCarlo & Johnson, 1999, 2000, 2002; Ghazanfar & Nicolelis, 1999). Any other sensory modality with the concept of a spatial receptive field/response area can be generalized to include the dimension of time and stands to gain from the methodology (Ghazanfar & Nicolelis, 2001; Linden & Schreiner, 2003).

Using spectro-temporally rich stimuli and systems analysis methods, STRFs of hundreds of units in AI in the anesthetized and awake ferret have been measured, mapped, and compared to those obtained from one and two-tone stimuli (Depireux,

Simon, Klein, & Shamma, 2001; Klein, Simon, Depireux, & Shamma, 2006). For these techniques to be useful requires that responses to such broadband stimuli have a *robustly* linear component, an assumption that has been investigated, and, in many types of cells, confirmed (Escabi & Schreiner, 2002; Klein, Simon, Depireux, & Shamma, 2006; Kowalski, Depireux, & Shamma, 1996; Schnupp, Mrsic-Flogel, & King, 2001; Shamma, Versnel, & Kowalski, 1995; Theunissen, Sen, & Doupe, 2000). Robustness requires, at least, that when the STRF is measured using stimulus sets with very different spectrotemporal profiles, the STRF is (approximately) independent of stimulus set (Klein, Simon, Depireux, & Shamma, 2006). The most important consequence of linearity, the superposition principle (Papoulis, 1987), requires that the responses to combinations of spectrotemporal envelopes be linearly additive and predictable from the STRF. These results were successfully applied to predictions of responses to spectra composed of multiple moving ripples (Klein, Simon, Depireux, & Shamma, 2006; Kowalski, Depireux, & Shamma, 1996). It should be noted that many researchers have also observed a significant proportion of units that are unpredictable or poorly responsive, and cannot be described by purely linear assumptions (Theunissen, Sen, & Doupe, 2000; Ulanovsky, Las, & Nelken, 2003). Nor does a response which is robustly linear disallow nonlinearities: neurons with a substantially linear component typically contain substantial nonlinearities as well, such as having firing rates above the mean with a much larger dynamic range than firing rates below the mean. In particular, response profiles typically have strong static non-linearities and yet their linear response is still robust (van Dijk, Wit, Segenhout, & Tubis, 1994).

In this work, we show that those neurons in AI that are well described by STRFs have a special property, which we call temporal symmetry. Temporal symmetry means that all temporal cross-sections of any STRF are the same time function (i.e. impulse response), except for a scaling and a "Hilbert rotation" (defined below). We further show that temporal symmetry has strong implications for the functional neural connectivity of neurons in AI, both in their thalamic input—from the ventral Medial Geniculate Body (MGB)—and in their intracortical inputs. In fact, most simple, otherwise compelling, models of functional neural connectivity of neurons in AI are disallowed physiologically because they violate the property of temporal symmetry. Other models, still biologically plausible, are suggested which obey the temporal symmetry property.

Most of the mathematical treatments discussed in this work arose from the context of linear systems. It is crucial, however, that the linear systems treatment itself lies in the context of the more general non-linear framework, e.g. of Volterra and Wiener (Eggermont, 1993; Rugh, 1981). Thus, even though the system has strong non-linearities in addition to its linearity, as long as the linear component of the overall response is robust, the estimated STRF itself should be robust. This robust STRF, with its property of temporal symmetry, can justifiably be used to strongly constrain models of neural connectivity in AI. To reiterate, the presence of strong non-linearities are consistent with the presence of a robust linear component, and that robust linear component (here, the STRF) places strong constraints on models of neural connectivity.

# *METHODS*

## Defining the Spectro-Temporal Response Field (STRF)

In the auditory system, the Spectro-Temporal Response Field (STRF) is a function of both time and frequency, $h(t, x)$, where $t$ is response time (e.g. in ms) and $x = \log_2(f/f_0)$ is the number of octaves above a reference frequency $f_0$. The firing rate of the neuron $r(t)$, has a linear component $r_{lin}(t)$ given by the linear, temporal convolution of its STRF, $h(t, x)$, with the spectro-temporal envelope of the stimulus $s(t, x)$:

$$\begin{aligned} r_{lin}(t) &= \int dt' \int dx\, s(t'-t, x) h(t', x) \\ &= \int dx\, s(t, x) *_t h(t, x) \end{aligned}$$

(1)

where $*_t$ means convolution in the $t$-dimension (but not $x$). The full firing rate $r(t)$ will differ from the linear rate $r_{lin}(t)$, to the extent that the system is not entirely linear, but the STRF determines all of the firing rate that is linear with respect to the spectro-temporal envelope of the stimulus (Depireux, Simon, & Shamma, 1998). Crucially, even if the system has strong non-linearities, so long as the linear properties are robust, $r_{lin}(t)$ will be consistently determined entirely by the STRF and the spectro-temporal envelope of the stimulus.

There are several straightforward interpretations of the STRF, all ultimately equivalent. Four are presented below: The first two interpret the two dimensional STRF as a collection of one-dimensional response profiles. The third interprets the entire STRF as the spectro-temporal representation of an optimal (acoustic) stimulus.

The fourth is a general qualitative scheme for predicting the response to any broadband stimulus from the STRF.

[Figure 2 about here.]

*Spectral Response Field Interpretation*: Any STRF cross-section at a single moment in time (i.e. a vertical cross-section) can be interpreted as an instantaneous spectral Response Field (Figure 2a, bottom left). In this interpretation, a peak in the Response Field indicates a high (instantaneous) spike rate when the stimulus has enhanced power in that spectral band. A dip in the Response Field indicates a low (instantaneous) spike rate when the stimulus has enhanced power in that spectral band (e.g. from side-band inhibition). A cross section can be examined at any instant in time, so the STRF can be interpreted as a time-evolving Response Field.

*Impulse Response Interpretation*: Any STRF cross-section at a single frequency (i.e. a horizontal cross-section) can be interpreted as a narrowband Impulse Response (Figure 2a, top right). In this interpretation, the Impulse Response is the response to the instantaneous presentation of high power in a narrow band. There is a separate Impulse Response for every frequency channel, so the STRF can be interpreted as a spectrally ordered collection of Impulse Responses.

*Optimal Stimulus Interpretation*: The entire STRF can be interpreted as a whole by flipping the time axis and interpreting the new image as proportional to the spectrogram of a stimulus. In this picture, stimulus time evolves (from left to right), growing less negative until it stops at $t = 0$ (Figure 2b). The stimulus is optimal in a very specific sense: of all stimuli with the same power, the (linear estimate of the) stimulus that gives the highest spike rate for that power is proportional to the time reversed STRF. This is a straightforward result from linear systems theory (see e.g. deCharms, Blake, & Merzenich, 1998; Papoulis, 1987).

*General Qualitative Interpretation*: The response of the entire neuron to any broadband stimulus can be estimated by convolving features of the stimulus spectrogram with features of the STRF (Figure 2c). Regions of the STRF that are positive ("excitatory") will contribute positively to the firing rate when the stimulus has enhanced power in that spectral band (enhanced relative to the background stimulus level). Similarly, regions of the STRF that are negative ("inhibitory") will contribute negatively to the firing rate when the stimulus has enhanced power in that spectral band. Naturally, regions of the STRF that are positive ("excitatory") will also contribute *negatively* to the firing rate when the stimulus has *diminished* power in that spectral band (diminished relative to the background stimulus level). Less intuitive but still natural, regions of the STRF that are negative ("inhibitory") will contribute *positively* to the firing rate when the stimulus has *diminished* power in that spectral band (since the tendency to reduce firing rate is itself weakened). The firing rate at any given moment is the sum of all these products, with the appropriate weighting. This last interpretation is really just a verbal description of Eq. 1.

## Measuring the Spectro-Temporal Response Field (STRF)

The STRF can be estimated in many different ways, but all are equivalent to inverting Eq. 1, i.e., cross-correlating the full neural response rate $r(t)$ with the spectro-temporal envelope of the stimulus $S(t,x)$. This is also known as spike-triggered averaging. Many types of stimuli can be used to measure an STRF, as long as they are sufficiently spectro-temporally rich. Among the stimuli used are auditory ripples (e.g. Depireux, Simon, & Shamma, 1998; Kowalski, Depireux, & Shamma, 1996; Miller & Schreiner, 2000; Qiu, Schreiner, & Escabi, 2003), auditory m-sequences (Kvale & Schreiner, 1997), random chords (e.g. deCharms, Blake, & Merzenich, 1998; Valentine & Eggermont, 2003), and spectro-temporally rich natural sounds (e.g. Theunissen, Sen, & Doupe, 2000).

## Relationship to Vision and the Spatio-Temporal Response Field

The visual system has neurons that are well characterized by the analogous quantity, the Spatio-Temporal Response Field, $h(t,\vec{x})$, whose arguments are the two-dimensional angular distance, $\vec{x}$ and time $t$, and whose temporal convolution with a Spatio-Temporal stimulus, $s(t,\vec{x})$, (e.g. drifting contrast gratings) gives the linear firing rate of the cell:

$$r_l(t) = \int d\vec{x}\, s(t,\vec{x}) *_t h(t,\vec{x}) \qquad (2)$$

which is the same as Eq. 1 but with retinotopic position $\vec{x}$ instead of cochleotopic position $x$.

All the methodology, above and below, relevant to Spectro-Temporal Response Fields $h(t,x)$ also applies to Spatio-Temporal Response Fields $h(t,\vec{x})$, with the following substitutions: $x \rightarrow \vec{x}$, $\Omega \rightarrow \vec{\Omega}$, and $\Omega x \rightarrow \vec{\Omega} \cdot \vec{x}$. The applications below will apply only to the extent that cortical visual processing is comparable to cortical auditory processing, and to their respective physiological properties, and of course which area within cortex is being characterized.

Stimuli used to calculate visual STRFs must have contrast that changes in both space and time. Typical stimuli range from drifting contrast gratings, randomly changing dots or bars, m-sequences, or more complex patterns (see, e.g. De Valois, Cottaris, Mahon, Elfar, & Wilson, 2000; De Valois & De Valois, 1988; Reid, Victor, & Shapley, 1997; Richmond, Optican, & Spitzer, 1990; Sutter, 1992; Victor, 1992). These spatiotemporally rich stimuli can be compared to the spectro-temporally rich auditory stimuli described above (auditory ripples, random chords, and spectro-temporally rich natural sounds).

We use the abbreviation STRF to apply to both spectral (auditory) and spatial (visual) cases. Context will make clear to which case it refers.

## Rank and Separability

The *rank* of a two dimensional function, such as an STRF or a Spectro-Temporal Modulation Transfer Function ($MTF_{ST}$), captures one aspect of how simple the function is. When a two dimensional function is the simple product of two one-dimensional functions, i. e. $h(t,x) = f(t)g(x)$, this captures an important notion of simplicity. When this occurs, the function is of rank 1. When the sum of two products is required, e.g. $h(t,x) = f_A(t)g_A(x) + f_B(t)g_B(x)$, the function is of rank 2. (In cases of rank 2 and higher, we demand that each temporal function $f_i(t)$ be linearly independent of every other temporal function $f_j(t)$, and the same for the spectral functions $g$; otherwise we could have used a smaller number of terms.) A rank 2 function is clearly not as simple as a rank 1 function, but nevertheless can be expressed rather concisely. In general, the *rank* of any two dimensional function is the minimum number of simple products needed to describe the function (When the functions are approximated as discrete, the definition of rank is identical to the definition of the algebraic rank of a matrix).

An STRF of rank 1, also called *fully separable*, can be written

$$h^{FS}(t,x) = f(t)g(x),$$
(3)

which has this simple interpretation: the temporal processing of the STRF is performed independently of the spectral processing (and of course vice versa). A simple model of a neuron with this property is that its spectral processing is due purely to inputs from presynaptic neurons with a range of center frequencies, while the temporal processing is due to integration in the soma of all inputs arriving from all dendrites. For many peripheral neurons, this model is a good one. An STRF of rank 1 is also called fully separable because its processing separates cleanly into independent spectral and temporal processing stages. The example STRF in Figure 1a is fully separable, and this can be verified by noting that all spectral (vertical) cross-sections have the same shape (the shape of the spectral function $g(x)$), differing only in amplitude (and possible sign). Similarly, all temporal (horizontal) cross-sections have the same shape (the shape of the temporal function $f(t)$), differing only in amplitude (and possibly sign).

An STRF of rank 2 is somewhat less simple and is somewhat less straightforward to interpret.

$$h^{R2}(t,x) = f_A(t)g_A(x) + f_B(t)g_B(x) \tag{4}$$

One interpretation comes from noting that this $h^{R2}(t,x)$ can be written as the sum of two fully separable STRFs, $h_A^{FS}(t,x) = f_A(t)g_A(x)$ and $h_B^{FS}(t,x) = f_B(t)g_B(x)$. This implies a possible, but less than satisfying, interpretation: the neuron has exactly two neural inputs, each of which has a fully separable STRF, and then simply adds them. Below we will present more realistic interpretations, consistent with known physiology. An STRF described by a generic two-dimensional function would not necessarily have the same physiologically motivated interpretations or models.

An STRF of general rank *N* can be written

$$h^{RN}(t,x) = \underbrace{f_A(t)g_A(x) + f_B(t)g_B(x) + \ldots + f_Z(t)g_Z(x)}_{N \text{ terms}}. \tag{5}$$

As the rank of an STRF increases, more and more complexity is permitted. Figure 1b demonstrates a simulated STRF of high rank (though still well localized in time and spectrum). STRFs of this complexity are not seen in AI (Klein, Simon, Depireux, & Shamma, 2006). In general, *higher rank* implies more STRF complexity. On the other hand, *lower rank* suggests there is a specific property (constraint) that causes this simplicity.

# Singular Value Decomposition Analysis of the STRF

Singular Value Decomposition (SVD) is a method that can be applied to any finite dimensional matrix (e.g. a discretized version of the STRF) to establish both its rank and a unique re-expression of the matrix as the sum of terms whose number is the rank of the matrix (Hansen, 1997; Press, Teukolsky, Vettering, & Flannery, 1986). The SVD decomposition of a matrix $M$ takes the form

$$M_{ij} = \underbrace{\Lambda_A u_{Ai} v_{Aj}^{\mathrm{T}} + \Lambda_B u_{Bi} v_{Bj}^{\mathrm{T}} + \ldots + \Lambda_Z u_{Zi} v_{Zj}^{\mathrm{T}}}_{N\,\text{terms}} \qquad (7)$$

where $N$ is the rank of the matrix, $u$ and $v$ are vectors normalized to have unit power, and each $\Lambda$ is the term's RMS power. If we discretize the STRF into a finite number of frequencies and time steps, $x = \{x_i\} = (x_1, \ldots, x_M)$ and $t = \{t_j\} = (t_1, \ldots, t_N)$, so that $h(t_i, x_j) = \{h_{ij}\} = h(t_1, \ldots, t_N; x_1, \ldots, x_M)$, we see that

$$h_{ij} = \underbrace{\Lambda_A u_A(x_i) v_A(t_j) + \Lambda_B u_B(x_i) v_B(t_j) + \ldots + \Lambda_Z u_Z(x_i) v_Z(t_j)}_{N\,\text{terms}} \qquad (8)$$

where $N$ is the rank of the STRF, the $u$ and $v$ vectors are normalized to have unit power, and each $\Lambda$ is the RMS power of its term. This is the same as Eq. 5, except that time and frequency have been discretized, and thus the STRF has been discretized also.

What makes SVD unique among decompositions is that 1) it automatically orders the terms by decreasing power: $\Lambda_A > \Lambda_B > \ldots > \Lambda_Z$; 2) each column $u_A, u_B, \ldots, u_Z$ is orthogonal to all the others; 3) each row $v_A^{\mathrm{T}}, v_B^{\mathrm{T}}, \ldots, v_Z^{\mathrm{T}}$ is orthogonal to all the others. The mathematical specifics are described well in textbooks (see, e.g., Press, Teukolsky, Vettering, & Flannery, 1986) and will not be covered here.

Mathematically, SVD is intimately related to Principle Component Analysis (PCA), and both are used for a variety of analytic purposes, including noise reduction (Hansen, 1997).

Since measured STRFs are made with noisy measurements (the noise arising from both neural variability and instrument noise), the true rank of the STRF must be estimated. There are a variety of methods to do this (G. W. Stewart, 1993) but they all use the same conceptual framework: once the power of the noise is estimated, then all SVD components with power greater than the noise can be considered signal, and the number of components satisfying this criterion is the estimate of the rank. This estimate of rank is biased (more noise results in a lower rank estimate), but it has been shown that for range of Signal-to-Noise ratios and the STRFs used in this study, noise is not an impediment to measuring high rank (Klein, Simon, Depireux, & Shamma, 2006).

SVD also motivates us to recast Eq. 5 into its continuous form

$$h^{RN}(t,x) = \underbrace{\Lambda_A v_A(t) u_A(x) + \Lambda_B v_B(t) u_B(x) + \ldots + \Lambda_Z v_Z(t) u_Z(x)}_{N \text{ terms}},\qquad(9)$$

where the $u$ and $v$ functions have unit power and each $\Lambda$ is the RMS power of its term. Compared to Eq. 5, it more complex, but it is less arbitrary: decompositions of the form of Eqs. 3, 4, and 5 are not unique since amplitude can be arbitrarily shifted between the temporal and spectral components. In Eq. 8, all amplitude information is explicitly shared within each term by each $\Lambda_i$ coefficient. When the technique of SVD, which is designed for discrete matrices, is applied to continuous two dimensional functions, as in the case of Eq. 8, it is called the Singular Value

Expansion (Hansen, 1997). We will go back and forth between the continuous and discretized versions of the STRF without loss of generality (so long as $N$ is finite), depending on which formalism is more beneficial.

## Hilbert Transform & Partial Hilbert Transforms/Rotations

We now discuss the Hilbert transform, a standard tool in signal processing, and necessary for the phenomenon of temporal symmetry.

The Hilbert transform of a function produces the same function but with all its phase components shifted by 90°. This can be seen in the Fourier domain. For a function $f(t)$ with Fourier transform $F(\omega)$, i.e.

$$F(\omega) = \mathcal{F}_\omega \big[ f(t) \big] = \int dt\, f(t) \mathrm{e}^{-j\omega t}$$
$$f(t) = \mathcal{F}_t^{-1} \big[ F(\omega) \big] = (2\pi)^{-1} \int d\omega\, F(\omega) \mathrm{e}^{j\omega t} \ , \tag{10}$$

the Hilbert transform, designated by $\mathcal{H}$ or $\hat{}$, is defined by

$$\hat{f}(t) = \mathcal{H}\big[ f(t) \big] = \mathcal{F}_t^{-1} \big[ \mathrm{sgn}(\omega) e^{j\pi/2} F(\omega) \big], \tag{11}$$

where $e^{j\pi/2} = j$ is a rotation by 90° in the complex plane (the role of $\mathrm{sgn}(\omega)$ guarantees that the Hilbert transform of a real function is itself a real function). This rotation of phase by 90° means that the Hilbert transform of any sine wave is a cosine wave, and the Hilbert transform of any cosine wave is the negative sine wave, but unlike differentiation, the amplitude is unchanged by the operation.

An important property of the Hilbert transform is that it is orthogonal to the original function, and yet it still has the same frequency content (aside from the DC component, i.e. its mean, which is zeroed out). $\hat{f}(t)$ is said to be "in quadrature" with $f(t)$; a demonstration is illustrated in Figure 3.

For the remainder of this section, we assume that any function $f(t)$ which will be Hilbert transformed has mean zero (or has had its mean subtracted manually).

[Figure 3 about here.]

The double application of a Hilbert transform, since applying two successive 90° rotations is equivalent to one 180° rotation, is just a sign inversion.

$$\mathcal{H}\big[\mathcal{H}\big[f(t)\big]\big] = \hat{\hat{f}}(t) = -f(t) \tag{12}$$

It is also useful to define a partial Hilbert transform. As pointed out above, a Hilbert transform of a function can be viewed as a 90° rotation in a mixing angle plane, so one can define a partial version of the transform

$$f^{\theta}(t) = \sin\theta \ \hat{f}(t) + \cos\theta \ f(t). \tag{13}$$

In this convention, note that $\hat{f}(t) = f^{\pi/2}(t)$, $f(t) = f^{0}(t)$, and $\hat{\hat{f}}(t) = f^{\pi}(t) = -f(t)$. Thus a partial Hilbert transform still has the same frequency content as the original function, but its phase "rotation" is not restricted to 90° and can be any angle on the complex plane.

Physiological examples of the Hilbert transform have been demonstrated in the visual system, and have been named "lagged" cells (De Valois, Cottaris, Mahon, Elfar, &

Wilson, 2000; Humphrey & Weller, 1988; Mastronarde, 1987a, 1987b). These lagged cells are located in the Lateral Geniculate Nucleus (LGN), one of the visual thalamic nuclei. We will continue this nomenclature and call any neuron whose impulse response is the Hilbert transform of another the lagged version of the latter. We will further generalize, and call any neuron whose impulse response is the partial Hilbert transform (Hilbert rotation) of another, the "partially lagged" version of the latter. Note that the lag is a phase lag, not a time lag.

The full and partial Hilbert transform/rotation is not restricted to the time domain and is equally applicable to the spectral domain. For example,

$$\mathcal{H}\big[g(x)\big] = \hat{g}(x) \tag{14}$$

$$\mathcal{H}\big[\mathcal{H}\big[g(x)\big]\big] = \hat{\hat{g}}(x) = -g(x) \tag{15}$$

$$g^{\theta}(x) = \sin\theta \; \hat{g}(x) + \cos\theta \; g(x). \tag{16}$$

## Temporal Symmetry

An important class of STRFs consists of those for which all temporal cross-sections (i.e. each cross section at a constant spectral index $x_c$), of the given STRF are related to each other by a simple scaling, $g$, and rotation, $\theta$, of the same time function:

$$h(t, x_c) = g_{x_c} f^{\theta_{x_c}}(t), \tag{17}$$

where each scaling and rotation can depend on $x_c$. Since this is then true for all spectral indices $x$, we call the system Temporally Symmetric and write it in the functional form

$$h^{TS}(t,x) = g(x)f^{\theta(x)}(t) \tag{18}$$

The meaning is still the same: all temporal cross-sections are related to each other by a simple scaling and rotation of the same time function. There is only one function of $t$, i.e. $f(t)$, and Hilbert rotations of it (demonstrated in Figure 4).

[Figure 4 about here.]

Using the definition of the Hilbert rotation Eq. 13, we can re-express Eq. 18 to explicitly show that a temporally symmetric STRF is rank 2 (i.e. is the sum of two linearly independent product terms):

$$\begin{aligned}
h^{TS}(t,x) &= g(x)f^{\theta(x)}(t) \\
&= g(x)\cos\theta(x)f(t) + g(x)\sin\theta(x)\hat{f}(t) \\
&= f(t)g_A(x) + \hat{f}(t)g_B(x)
\end{aligned} \tag{19}$$

where

$$\begin{aligned}
g_A(x) &= g(x)\cos\theta(x) \\
g_B(x) &= g(x)\sin\theta(x) \\
\tan\theta(x) &= \frac{g_B(x)}{g_A(x)} \\
g^2(x) &= g_A^2(x) + g_B^2(x).
\end{aligned} \tag{20}$$

We will often use the form of Eq. 19, which is completely equivalent to Eq. 18. In Eq. 19 it is explicit that a temporally symmetric STRF has rank 2 and cannot have higher rank.

For systems, which are not exactly temporally symmetric but are of rank 2, or for systems, which have been truncated by SVD to rank 2, we can define an index of temporal symmetry, $\eta_t$. This index ranges from 0 to 1, where $\eta_t = 1$ for the temporally symmetric case and $\eta_t = 0$ when the two time functions are temporally unrelated. First we put Eq. 19, which is explicitly rank 2, into the form of Eq. 8:

$$h^{R2}(t,x) = \Lambda_A v_A(t) u_A(x) + \Lambda_B v_B(t) u_B(x).$$  (21)

Since the $u$ and $v$ functions have unit power, we define the index of temporal symmetry to be the magnitude of the normalized complex inner product between the two temporal analytic signals (Cohen, 1995)

$$\eta_t = \left| \int \tfrac{1}{2} \big( v_A(t) + j\hat{v}_A(t) \big)^* \big( v_B(t) + j\hat{v}_B(t) \big) dt \right|,$$  (22)

where $^*$ is the complex conjugate operator. The rank 1 case, since it is automatically temporally symmetric, is also given the value $\eta_t = 1$.

Temporal symmetry's cousin, spectral symmetry can be defined analogously.

$$\begin{aligned}
h^{SS}(t,x) &= f(t) g^{\theta(t)}(x) \\
&= f(t)\cos\theta(t) g(x) + f(t)\sin\theta(t) \hat{g}(x) \\
&= f_A(t) g(x) + f_B(t) \hat{g}(x)
\end{aligned}$$  (23)

where

$$\begin{aligned}
f_A(t) &= f(t)\cos\theta(t) \\
f_B(t) &= f(t)\sin\theta(t) \\
\tan\theta(t) &= \frac{f_B(t)}{f_A(t)} \\
f^2(t) &= f_A^2(t) + f_B^2(t).
\end{aligned}$$  (24)

and

$$\eta_s = \left| \int \tfrac{1}{2} \big( u_A(x) + j\hat{u}_A(x) \big)^* \big( u_B(x) + j\hat{u}_B(x) \big) dx \right|.$$  (25)

## Spectro-Temporal Modulation Transfer Functions (MTF$_{ST}$)

Just as any STRF may have the property of temporal or spectral symmetry, it may also have the property of quadrant separability. Quadrant separability is most easily described in terms of the Spectro-Temporal Modulation Transfer Function (MTF$_{ST}$), which is presented here.

The STRF, which is two-dimensional, can also be represented by its two-dimensional Fourier transform, or its closely related partner, the MTF$_{ST}$

$$H(w, \Omega) = \mathcal{F}_{w, \Omega} \big[ h(t, -x) \big]$$
$$= \int dt \int dx \, h(t, x) e^{2\pi j(-wt + \Omega x)} \tag{26}$$

where $w$ and $\Omega$ are the coordinates Fourier-conjugate to $t$ and $x$ respectively (see Depireux, Simon, & Shamma, 1998 for sign conventions). Examples are shown in Figure 5.

[Figure 5 about here.]

It follows that the inverse Fourier transform of $H(w, \Omega)$ gives the STRF of the cell.

$$h(t, x) = \mathcal{F}_{t, -x}^{-1} \big[ H(w, \Omega) \big] \tag{27}$$

The MTF$_{ST}$ is a Fourier transform of the STRF, and is also used to characterize auditory processing. E.g. to the extent that the STRF represents a stimulus that the neuron prefers, the Fourier transform provides an analytical description of the features of that stimulus. Power at low $w$, which has dimension of cycles/s or Hz, corresponds to smoother temporal features, or slower temporal evolution. Power at high $w$ corresponds to finer temporal features and faster temporal resolution. Lower (vs. higher) $\Omega$, which has dimensions of cycles/octave, corresponds to smoother (vs. finer) scale spectral features, e.g. broad (vs. sharp) peaks, formants (vs. harmonics), etc..

The four possible sign combinations of $w$ and $\Omega$ break the MTF$_{ST}$ into 4 quadrants, numbered 1 ($w, \Omega > 0$), 2 ($w < 0, \Omega > 0$), 3 ($w, \Omega < 0$), and 4 ($w > 0, \Omega < 0$). From Eq. 26 it can be seen that $H(w, \Omega)$ is a complex valued function. Because $h(t, x)$ is purely real, the MTF$_{ST}$ has a complex-conjugate symmetry,

$$H(-w, -\Omega) = H^*(w, \Omega).$$ (28)

Eq. 28 also holds for the Fourier Transform of any real function of $t$ and $x$. This means that the value of the MTF$_{ST}$ at any point in quadrant 3 is fully determined by the value at the reflected point in quadrant 1 (and similarly for the pair quadrant 4 and quadrant 2).

The MTF$_{ST}$, its properties and interpretations, are discussed in greater detail elsewhere (Depireux, Simon, & Shamma, 1998; Klein, Simon, Depireux, & Shamma, 2006).

## Directionality and Quadrant Separability

When the STRF is separable, the MTF$_{ST}$ is separable, because the Fourier transform of the STRF is given by the simple products of the Fourier transforms of $f(t)$ and $g(x)$. It was noticed that even when the STRF is not separable, the quadrants of the MTF$_{ST}$ are still individually separable (Klein, Simon, Depireux, & Shamma, 2006; Kowalski, Depireux, & Shamma, 1996), but neither the significance nor the origin of this property was well understood. See McLean and Palmer (1994) for the analogous case in vision. With the discovery of temporal symmetry and the relationship between temporal symmetry and quadrant separability, the significance is now clear, as will be shown.

One of the most useful properties of the Fourier representation (i.e. the MTF$_{ST}$) over the spectro-temporal representation (i.e. the STRF) is that in the Fourier representation, the contributions of the response, to stimuli with upward and downward moving spectral features, are explicitly segregated in different quadrants. The response to any downward moving component is governed entirely by the MTF$_{ST}$ in quadrant 1, and the response to any upward moving component is governed entirely by the MTF$_{ST}$ in quadrant 2.

*Quadrant Separability* is a particular generalized symmetry property that an STRF and its MTF$_{ST}$ may have, but it is only obvious when seen in the MTF$_{ST}$ domain: within each quadrant, the MTF$_{ST}$ is separable. For example in quadrant 1, where both $w$ and $\Omega$ are positive, the MTF$_{ST}$ is the simple product of a horizontal (temporal) function and a vertical (spectral) function. Similarly, in quadrant 2, where $w$ is negative and $\Omega$ is positive, the MTF$_{ST}$ is the simple product of a different horizontal (temporal) function and a different vertical (spectral) function. An example is shown in Figure 5c. Quadrant separable $MTF_{ST}$s are defined and characterized with more mathematical detail in the appendix.

Historically in audition, the property of quadrant separability was noticed when the spectro-temporal modulation transfer function measured in quadrant 1 was separable, and the spectro-temporal modulation transfer function measured in quadrant 2 was also separable, but the two separable functions were not the same (Depireux, Simon, Klein, & Shamma, 2001; Kowalski, Depireux, & Shamma, 1996) (if it is separable in quadrants 1 and 2, then it is automatically separable in quadrants 3 and 4, from Eq. 28). In vision studies, quadrant separability was invoked for the notion of directional selectivity (Watson & Ahumada, 1985). In this work we argue that quadrant separability in the auditory system is due entirely to temporal symmetry.

Quadrant separability is a property of both the STRF and the $MTF_{ST}$, though visible only in the $MTF_{ST}$. Nevertheless, since the STRF and $MTF_{ST}$ are just different representations of the same response properties, it is a property that is held (or not) by both. It is shown in the appendix that a quadrant separable STRF can always be written in the form

$$h^{QS}(t,x) = f_A(t)g_A(x) + \hat{f}_B(t)\hat{g}_B(x) + \hat{f}_A(t)g_B(x) + f_B(t)\hat{g}_A(x) \qquad (29)$$

which is shown in the Appendix to be of rank 4, unless additional symmetries, such as those discussed later, reduce the rank to 2 or 1.

## Quadrant Separability and Temporal Symmetry

Comparing Eq. 29 to Eq. 19 we can see that the temporally symmetric $h^{TS}(t, x)$ has the same form as $h^{QS}(t, x)$ for the special case that $f_B(t) = 0$. Thus, *a temporally symmetric STRF is automatically quadrant separable*. It is not a generic quadrant separable STRF, since its rank is not 4 but 2 (by inspection of Eq. 19). It nevertheless possesses the defining property of quadrant separability: each quadrant of its MTF$_{ST}$ is separately separable (e.g. Figure 5c). We will use this property below to show that an STRF that is not quadrant separable cannot be temporally symmetric.

## Quadrant Separability and General Symmetries

There are three ways of taking the generic quadrant separable STRF of rank 4 and finding a generalized symmetry which causes it to be lower rank.

The generalized symmetry of temporal Hilbert symmetry has already been discussed. It can be seen by taking the most general form of a quadrant separable STRF, Eq. 29, and noting that setting $f_B(t) = 0 = \hat{f}_B(t)$, so that only $f_A(t)$ and $\hat{f}_A(t)$ survive as temporal functions, reduces the number of independent components (i.e. the rank) from 4 to 2.

The generalized symmetry of spectral symmetry works analogously, since the mathematics is blind to the difference between time and spectrum. It can be seen by noting that setting $g_B(x) = 0 = \hat{g}_B(x)$, so that only $g_A(x)$ and $\hat{g}_A(x)$ survive as spectral functions, also reduces the number of independent components (i.e. the rank) from 4 to 2. This gives the spectrally symmetric, quadrant separable STRF

$$h^{ss}(t,x) = f_A(t)g_A(x) + f_B(t)\hat{g}_A(x).$$ (30)

We are not aware of any physiological system in which this generalized symmetry is realized.

Finally, a third generalized symmetry is pure *Directional Selectivity*. In the case of this generalized symmetry, we set

$$f_A(t) = \hat{f}_B(t)$$
$$g_A(x) = \hat{g}_B(x)$$ (31)

which, when combined with the identity that the double application of the Hilbert operator is a Hilbert rotation of 180° and so is equivalent to multiplication by –1, gives

$$h^{DS}(t,x) = 2\left(f_A(t)g_A(x) - \hat{f}_A(t)\hat{g}_A(x)\right).$$ (32)

This is the case much discussed in the vision literature when STRF is interpreted as the visual Spatio-Temporal Response Field: both temporal and spectral functions are added in quadrature (Adelson & Bergen, 1985; Barlow & Levick, 1965; Borst & Egelhaaf, 1989; Chance, Nelson, & Abbott, 1998; De Valois, Cottaris, Mahon, Elfar, & Wilson, 2000; Emerson & Gerstein, 1977; Heeger, 1993; Maex & Orban, 1996; McLean & Palmer, 1994; A. T. Smith, Snowden, & Milne, 1994; Suarez, Koch, & Douglas, 1995; Watson & Ahumada, 1985). The result is a purely directionally selective response field, and again is rank 2.

These three symmetries reduce the rank of a quadrant separable spectro-temporal modulation transfer function from 4 to 2. In the appendix it is proven these are the only symmetries that can reduce the rank to 2, and that a quadrant separable STRF can never be of rank 3 (the rank 1 case is fully separable).

We are not aware of any physiological system which possesses a generic quadrant separable STRF, i.e. of rank 4.

## Counter-examples

After the preceding examples, one might be tempted to believe that all rank 2 STRFs are also quadrant separable, but this can be shown false with a simple counter example.

The form of Eq. 29 demonstrates how difficult it is to make a quadrant separable transfer function by combining fully separable inputs: half of the terms are required to be very specific functionals of the other half. Any departure leads to total inseparability.

For example, an STRF that is the linear sum of two separable STRFS

$$h^{R2}(t,x) = h_A^{FS}(t,x) + h_B^{FS}(t,x) \tag{33}$$

has a spectro-temporal modulation transfer function that is the linear sum of two separable spectro-temporal modulation transfer functions.

$$H^{R2}(w,\Omega) = H_A^{R2}(w,\Omega) + H_B^{R2}(w,\Omega) \tag{34}$$

This is not, in general, the form of a quadrant separable spectro-temporal modulation transfer function. As an example, the sum of two (almost) identical, fully separable STRFs whose only difference is that one is translated spectrally and temporally with respect to the other, gives an STRF which is strongly velocity selective and is not quadrant separable. This is demonstrated in Figure 6b, where the $MTF_{ST}$ clearly not quadrant separable. This proves that the STRF in Figure 6a cannot be temporally symmetric.

[Figure 6 about here.]

It is also simple to show an example of an STRF which is the sum of two temporally symmetric STRFs but which itself is not temporally symmetric. As an example, the sum of two (almost) identical, temporally symmetric STRFs whose only difference is that one is translated spectrally and temporally with respect to the other, gives an STRF which has rank 4, not the rank of 2 required by temporally symmetry. This is demonstrated in Figure 6c and 6d.

## Surgery and animal preparation

Data were collected from a total of 11 domestic ferrets (*Mustela putorius*) supplied by Marshall Farms (Rochester, NY). Eight of these ferrets were anesthetized during recording, and details of the surgery in full procedural details are as (Shamma, Fleshman, Wiser, & Versnel, 1993). These ferrets were anesthetized with sodium pentobarbital (40 mg/kg) and maintained under deep anesthesia during the surgery. Once the recording session started, a combination of Ketamine (8 mg/Kg/Hr), Xylazine (1.6 mg/Kg/Hr), Atropine (10 μg/Kg/Hr) and Dexamethasone (40 μg/Kg/Hr) was given throughout the experiment by continuous intravenous infusion, together with Dextrose, 5% in Ringer solution, at a rate of 1 cc/Kg/Hr, to maintain metabolic stability. The ectosylvian gyrus, which includes the primary auditory cortex, was exposed by craniotomy and the dura was reflected. The contralateral ear canal was exposed and partly resected, and a cone-shaped speculum containing a miniature speaker (Sony MDR-E464) was sutured to the meatal stump. The remaining three ferrets were used for awake recordings, with full surgical procedural details in (Fritz, Shamma, Elhilali, & Klein, 2003). In these experiments, ferrets were habituated to lie calmly in a restraining tube for periods of up to 4-6 hours. A head-post was surgically implanted on the ferret's skull (anesthetized with sodium pentobarbital (40 mg/kg) and maintained under deep anesthesia during the surgery), and used to hold the animal's head in a stable position during the daily neurophysiological recoding sessions. All experimental procedures were approved by the University of Maryland Animal Care and Use Committee and were in accord with NIH Guidelines.

## Recordings, spike sorting, and selection criteria

Action potentials from single units were recorded using tungsten microelectrodes with 5–7 MΩ tip impedances at 1 kHz. In each animal, electrode penetrations were made orthogonal to the cortical surface. In each penetration, cells were typically isolated at depths of 350–600 μm corresponding to cortical layers III and IV (Shamma, Fleshman, Wiser, & Versnel, 1993). In four anesthetized animals, neural signals were fed through a window discriminator and the time of spike occurrence relative to stimulus delivery was stored using a computer. In the other seven animals, the original neural electrical signals were stored for further processing offline. Using MATLAB software designed in-house, action potentials were then manually classified as belonging to one or more single units, and the spike times for each unit were recorded. The action potentials assigned to a single class met the following criteria: (1) the peaks of the spike waveforms exceeded 4 times the standard deviation of the entire recording; (2) each spike waveform was less than 2 ms in duration and consisted of a clear positive deflection followed immediately by a negative deflection; (3) the spike waveform classes were not visibly different from each other in amplitude, shape, or time course; (4) the histogram of inter-spike-intervals evidenced a minimum time between spikes (refractory period) of at least 1 ms; (5) the spike activity persisted throughout the recording session. This procedure occasionally produced units with very low spike counts. After consulting the distribution of spike counts for all units, units that fired less than half a spike per second were excluded from further analysis since a neuron with such a low spike rate requires longer stimulus durations to analyze.

## Stimuli and STRF measurement

The stimuli used were Temporally Orthogonal Ripple Combinations (TORCs), as described by Klein et al. (2006). TORCs are more complex than individually presented dynamic ripples, which are instances of band-passed noise whose spectral and temporal envelopes are co-sinusoidal, and can be thought of as auditory analogs of drifting contrast gratings used in vision studies (Shamma & Versnel, 1995; Shamma, Versnel, & Kowalski, 1995). The spectro-temporal envelope of a TORC is composed of sums of the spectro-temporal envelopes of temporally-orthogonal dynamic ripples. Temporally orthogonal means that no two ripple components of a given stimulus share the same temporal modulation rate (their temporal correlation is zero); therefore, each component evokes a different frequency in the linear portion of the response. Each TORC spectro-temporal envelope is composed from six dynamic ripples having the same spectral density $\Omega$ (in cyc/oct), but different $w$ spanning the range of 4 to 24 Hz. In the reverse-correlation operation, the 4 Hz response component is orthogonal to all stimulus components besides the 4 Hz ripple, the 8 Hz component is correlated only with the 8 Hz ripple, and so on. 15 distinct TORC envelopes are presented, with spectral density $\Omega$ ranging from –1.4 cycles/octave to +1.4 cycles/octave in steps of 0.2 cycles/octave. Each of those 15 TORCS is then presented again but with the reverse polarity of its spectro-temporal envelope (the "inverse-repeat method") to remove systematic errors due to even-order nonlinearities (Klein, Simon, Depireux, & Shamma, 2006; Moller, 1977; Wickesberg & Geisler, 1984). Multiple sweeps were presented for each stimulus. Sweeps of different stimuli, separated by 3 - 6 s of silence, were presented in a pseudorandom order, until a neuron was exposed to between 55 and 110 periods (13.75 - 27.5 s) of each stimulus. All stimuli had an 8 ms rise/fall time. All stimuli were gated and fed through an equalizer into the earphone. Calibration of the sound delivery system (to obtain a flat frequency response up to 20 kHz) was performed in situ with the use of a 1/8 in Brüel & Kjaer 4170 probe microphone. In the anesthetized case, the microphone was inserted into the ear canal through the wall of the speculum to within 5 mm of the tympanic membrane; the speculum and microphone setup resembles closely that suggested by Evans (1979). In the awake case the stimuli were delivered through inserted earphones that were calibrated *in situ* at the beginning of each experiment.

STRFs were measured by reverse correlation, i.e. spike triggered averaging (Klein, Depireux, Simon, & Shamma, 2000; Klein, Simon, Depireux, & Shamma, 2006). In particular, only the sustained portions of the responses were analyzed, since the first 250 ms interval of post-stimulus onset response was not used. To ensure reliable estimates, neurons with STRFs whose estimated Signal-to-Noise-Ratio was worse than 2 were excluded (Klein, Simon, Depireux, & Shamma, 2006).

## RESULTS

### *Temporal properties of the STRF*

In an arbitrary STRF, the spectral and temporal dimensions of the response are not necessarily related in any way. For example, the temporal cross-sections (impulse responses) at different frequencies ($x$) need not be systematically related to each other in any specific manner. However, Figure 7 illustrates an unanticipated result that we found to be prevalent in our data: All temporal cross-sections of a given STRF are related to each other by a simple scaling and rotation of the same time function. For example, if we designate the impulse response at the best frequency (1.2 kHz) to be the function $f(t) = f^{\theta=0}(t) = f^0(t)$, then the cross-section at 0.72 kHz is approximately its scaled *inverse* ($\approx -0.50 f(t) = 0.50 f^\pi(t)$); at 0.92 kHz, it is the scaled and *lagged* version ($\approx 0.33 f^{\pi/2}(t)$)). A schematic depiction of this STRF response property is shown in Figure 7. This property was defined above, in Eq. 18, to be "Temporal Symmetry." In the following section, we quantify and demonstrate the existence of this response property in almost all of our neurons.

[Figure 7 about here.]

In general, a *pure* scaling (i.e., no rotation, $\theta = 0$) among the temporal cross-sections of the STRF is expected if the STRF is fully-separable, that is it can be decomposed into *one* product of a temporal and a spectral function: $h^{FS}(t, x) = f(t)g(x)$. However, only 50% of our cells can be considered fully separable in the awake population, and 67% in the anesthetized population; the remainder all not fully separable. Consequently, this highly constraining and ubiquitous relationship involving a scaling *and* a rotation, described mathematically by Eq. 18

$$h^{TS}(t, x) = g(x)f^{\theta(x)}(t) \tag{18}$$

must imply another basic characteristic of cell responses as we discuss next.

## Rank and Temporal Symmetry in AI

To examine spectro-temporal interactions we applied SVD analysis to all STRFs derived from AI neurons in our experiments (see Klein, Simon, Depireux, & Shamma, 2006, for method), with results tabulated in Table 1. In the awake recordings, the STRF rank was found to be best approximated as rank 1 or 2 for 98% of the neurons; in the anesthetized case 97%. That is, the STRF is of the form of either Eq. 3 or Eq. 4. In general, an STRF need not be of such a low rank at all. For example, Figure 1b shows an otherwise plausible simulated high rank STRF. Does low rank reflect a simplicity to the underlying neural circuitry?

[Table 1 about here.]

In the awake recordings, STRFs of rank 1 constituted 50% of all neurons; in the anesthetized recordings, STRFs of rank 1 constituted 67% of all neurons. These STRFs are fully separable, and hence all temporal cross-sections of a given STRF are automatically related by a simple scaling. For rank 2 STRFs (awake 48%, anesthetized 30%), however, there is no such relationship. In fact, for a rank 2 STRF expressed as Eq. 4,

$$h^{R2}(t,x) = f_A(t)g_A(x) + f_B(t)g_B(x) \qquad (4)$$

there is no mathematical need for any particular relationship between its temporal cross-sections, but physiological evidence provides some.

[Figure 8 about here.]

The experimental results, for neurons with STRF of rank 2, in Figure 8a highlight a strong relationship between the temporal functions $f_A(t)$ and $f_B(t)$ isolated by the SVD analysis in Eq. 4, and compared via the temporal symmetry index defined in Eq. 22. A temporal symmetry index near 1 means that $f_A(t)$ and $f_B(t)$ are not arbitrary, but instead are closely related by a Hilbert transform. Comparing Eq. 4 and the last line of Eq. 19, we see that we must have:

$$f_B(t) = \hat{f}_A(t) \qquad (35)$$

Temporal symmetry in an STRF is an extremely restrictive property. SVD guarantees that $f_B(t)$ be orthogonal to $f_A(t)$, but does not restrict its frequency content in any substantive way (G. W. Stewart, 1990; G.W. Stewart, 1991; G. W. Stewart, 1993).

$f_B(t) = \hat{f}_A(t)$ is special since $\hat{f}_A(t)$ is the only function orthogonal to $f_A(t)$ which has the *same frequency content* as $f_A(t)$. In this sense, there is only *one* time function in the STRF, the one characterized by $f_A(t)$. This is not the case in the spectral dimension, where there is no single special spectral function picked out: $g_A(x)$ and $g_B(x)$ are mathematically and physiologically unconstrained: the population distribution for the analogous spectral symmetry index shown in Figure 8b shows no such close relationship. There is no evidence for spectral symmetry. Note also that fully separable (rank 1) neurons are not included in the population shown in Figure 8 since they are automatically temporally symmetric and spectrally symmetric.

Alternatively, we can begin with the STRF in the form of Eq. 10a and explain why the temporal cross-sections in our STRFs exhibit the specific relationship depicted earlier in Figure 7. The impulse response at any $x$ can be thought of as a linear combination of $f_A(t)$ and $\hat{f}_A(t)$, which by Eq. 19 always gives a scaled version of a Hilbert rotated $f_A(t)$.

Another test for the presence of temporal symmetry arises from comparing STRFs approximated by two different means: the first two terms of the singular value expansion, versus the first term of the quadrant separable expansion in singular values (see Klein, Simon, Depireux, & Shamma, 2006 for details). The former ("rank 2 truncation") is of rank 2 by construction. The latter ("quadrant separable truncation"), since it is quadrant separable by construction, should be of rank 4 unless the STRF has a symmetry. Since there is no mathematical reason they should give the same result, the near unity correlation coefficient between the two shown in Figure 9a is experimental evidence that they are identical up to measurement error: the quadrant separable STRF is actually of rank 2 and therefore possesses a symmetry. The only quadrant separable STRFs with rank less then 4 must be temporally symmetric, spectrally symmetric, or directionally selective. The results shown in Figure 8b rule out spectral symmetry, and the analogous analysis for the index of directional selectivity (not shown) rules out directional selectivity. Therefore, STRFs' symmetry must be temporal symmetry.

For comparison, Figure 9b shows two other distributions. On the right is the same distribution as above but for rank 4 truncations instead of rank 2. The correlations *decrease*, indicating that rank 2 (and hence temporal symmetry) is a better estimate than rank 4 (generic quadrant separable STRFs are of rank 4, and only symmetric STRFs are of rank 2). The change must be small, since SVD orders contributions to the rank by decreasing power, but it need not have been negative. In the center of Figure 9b is the distribution of the same quantity as in Figure 9a, but with the STRFs permuted: the rank 2 truncation of the STRF is correlated with the quadrant separable truncation of every *other* STRF, for every rank 2 truncation. This distribution is broadly peaked around 0 (with the population scale normalized to be the same as in Figure 9a). This demonstrates that the skewness of the population toward unity in Figure 9a is not due to potentially confounding factors, such as the STRFs dominant power in the first ~100 ms, or having rank 2, or being quadrant separable.

[Figure 9 about here.]

Thus the results shown in Figure 8 and 9 provide evidence that that all these rank 2 neurons are indeed temporally symmetric. However, temporal symmetry is highly restrictive. There are many ways to obtain a rank 2 STRF and only a very small subset is temporally symmetric—yet almost all of AI rank 2 STRFs are! This finding must be a consequence of a fundamental anatomical and physiological constraint on the way AI units are driven by the spectro-temporally dynamic stimuli. In the remainder of this article, we demonstrate the simplest possible explanations that can give rise to these observed temporal symmetry properties.

## Implications & Interpretation of Temporal Symmetry for Neural Connectivity and Thalamic Inputs to AI

Neurons in AI receive thalamic inputs from ventral MGB, via layers III and IV (see Read, Winer, & Schreiner, 2002 for a recent review; P. H. Smith & Populin, 2001). In this section, we analyze the effects of the constraints of temporal symmetry on the temporal and spectral components of the thalamic inputs to an AI neuron. We are not analyzing the cells with fully separable STRFs directly: but it will be shown below that their analysis proceeds almost identically. Note again that the linear equations used in this analysis do not assume that all processing of inputs is linear; rather they assume that the linear component of the processing is strong and robust, in addition to all non-linear components of the processing.

The analysis below shows that there are physiologically reasonable models consistent with temporally symmetric neurons in, and throughout, AI. The models presented here require two features: that the STRFs of the thalamic inputs be fully separable and that some of the thalamic inputs be lagged (phase shifted), whether at the output of the thalamic neurons themselves or at their synapses onto AI neurons. Ventral MGB neurons possess STRFs which are consistent with being fully separable (Miller, Escabi, Read, & Schreiner, 2002; Miller, Escabi, & Schreiner, 2001; L. Miller, personal communication), but there has been no systematic study. Lagged neurons have not been reported in ventral MGB, though they exist in *visual* thalamus (Saul & Humphrey, 1990), and other lagging mechanisms that may be present in the auditory system are discussed below.

Alternatively, is very difficult to construct physiologically reasonable models consistent with temporal symmetry neurons in AI without these two features. We know of no such models, and we were not able to construct any. From this, we are forced to predict that these two features will be found. Independently of whether this occurs, however, some explanation is still needed for the temporal symmetry displayed strongly in AI, and in this section, we provide a reasonable basis.

## Simplistic Model of Thalamic Inputs

The last line of Eq. 19:

$$h^{TS}(t,x) = f_A(t)g_A(x) + \hat{f}_A(t)g_B(x).$$  (36)

explicitly demonstrates that the temporally symmetric STRF is rank 2. A simplistic interpretation of Eq. 36 is that a cell with a temporally symmetric STRF has two fully separable inputs, e.g. two cells in ventral MGB. Each of those two input cells has the same temporal processing behavior, except that one, whose temporal processing is characterized by $\hat{f}_A(t)$, is the Hilbert transform of (lagged with respect to) the other, with temporal processing characterized by $f_A(t)$. Mathematically, $\hat{f}(t)$ is in quadrature with $f(t)$. The two inputs may have different spectral distribution of their own inputs and/or different synaptic weights as inputs to the cortical cell: $g_A(x) \neq g_B(x)$. This is shown schematically in Figure 10a. (It is also consistent with this model that $g_A(x)$ could equal $g_B(x)$, but such a case reduces to the simple instance of a fully separable STRF.)

[Figure 10 about here.]

This interpretation is mathematically concise and explicitly demonstrates that the temporally symmetric STRF is rank 2. This is the maximally reduced form, and one may proceed to expand the form to demonstrate other forms that its inputs are permitted to take, e.g. allowing many inputs from thalamic or even intracortical connections. The goal of this section is to relax the strict decomposition implied by Eq. 36 (and imposed arbitrarily by SVD) and to rewrite it in a form that allows for physiologically reasonable inputs and physiologically reasonable somatic processing, and to analyze the restrictions that are imposed on those inputs. The more realistic decompositions below will allow us to reasonably model the thalamic inputs and the somatic processing by identifying them with terms in the decomposition, in a substantially more realistic interpretation than those simple interpretations presented above. The more realistic decompositions have an additional role as well: they contrast with models/decompositions which, appearing physiologically reasonable otherwise, conflict with the data and so can now be ruled out.

## Generalized Simplistic Model of Thalamic Inputs

First, the severity of the full Hilbert transform can be relaxed. The model decomposition can use a partial Hilbert transform $f^\theta(t)$ (see Eq. 13) instead of the full Hilbert transform. In Eq. 37 the first term of the decomposition has some temporal impulse response $f_A(t)$ and an arbitrary spectral response field $g_C(x)$. The second term has an impulse response $f_A^\theta(t)$ that is some Hilbert rotation of the first impulse response and an arbitrary spectral response field $g_D(x)$.

$$h^{TS}(t,x) = f_A(t)g_C(x) + f_A^\theta(t)g_D(x) \qquad (37)$$

which is equivalent to the previous decomposition in Eq. 36:

$$
\begin{aligned}
h(t,x) &= f_A(t)g_C(x) + f_A^\theta(t)g_D(x) \\
&= f_A(t)g_C(x) + \left(\sin\theta\,\hat{f}_A(t) + \cos\theta\,f_A(t)\right)g_D(x) \\
&= f_A(t)\underbrace{\left(g_C(x) + \cos\theta\,g_D(x)\right)}_{g_A(x)} + \hat{f}_A(t)\underbrace{\sin\theta\,g_D(x)}_{g_B(x)} \\
&= f_A(t)g_A(x) + \hat{f}_A(t)g_B(x) \\
&= h^{TS}(t,x)
\end{aligned}
\qquad (38)
$$

for any $\theta$ different from zero. This is physiologically more relevant since a partial ($\theta < 90°$) Hilbert transform may be a simpler operation for a neural process to perform than a full transform. The physiological interpretation of the first line of Eq. 38 is that the two temporal functions of the independent input classes need not be related by a full Hilbert transform (fully lagged cell); it is sufficient that one be a Hilbert "rotation" of the other (partially lagged cell). This is shown schematically in Figure 10b.

It should be noted that *exact* Hilbert rotations, for any $\theta \neq 0$ are ruled out by causality: it can be shown (e.g. Papoulis, 1987) that the Hilbert transform of a causal filter is acausal. Therefore, we would never expect an exact Hilbert transform (or exact temporal quadrature), only an approximate Hilbert transform. This is exemplified in the visual system where the Hilbert transform behavior of lagged cells is only a good approximation for the frequency band 1 – 16 Hz (Saul & Humphrey, 1990). In the auditory system we expect similar behavior: the (partial or full) Hilbert rotation will only be performed accurately in the relevant frequency band.

To restate, AI Models which attempt a decomposition of the form

$$h(t,x) = f_A(t)g_C(x) + f_B(t)g_D(x) \tag{39}$$

where $f_B(t) \neq f_A^\theta(t)$ for some $\theta$ are ruled out. This is because they violate the temporal symmetry property found in our STRFs. An example of a rank 2 STRF that is of the form of Eq. 39 but $f_B(t) \neq f_A^\theta(t)$ was shown Figure 6a. It is not temporally symmetric and is therefore ruled out as a model.

## Multiple Input Model

Continuing to add more physiological realism to the decomposition, we can allow for each of the two pathways to be made of multiple inputs, as long as their temporal structure is related, shown schematically in Figure 10c. In Eq. 40 the first sum in the decomposition is made up of a number of individual input components ($m = 1 \ldots M$) that all have the same temporal impulse response $f_A(t)$ and but may have individual spectral responses fields $g_{C_m}(x)$. The second sum is made up of a number of individual input components ($n = 1 \ldots N$) that each may have different temporal impulse responses $f_A^{\theta_n}(t)$, but all are given by some individual Hilbert rotation ($\theta_n$) of the initial impulse response, and individual spectral response fields $g_{D_n}(x)$.

$$h^{TS}(t,x) = \sum_{m=1}^{M} \left( f_A(t) g_{C_m}(x) \right) + \sum_{n=1}^{N} \left( f_A^{\theta_n}(t) g_{D_n}(x) \right) \qquad (40)$$

This decomposition is leading towards one based on the many neural inputs (i.e. $N + M$, which may be a very large number), each of which may have different spectral response fields and different Hilbert rotations, but must be related in their temporal structure. This is equivalent to the previous decomposition or to that in Eq. 36:

$$
\begin{aligned}
h(t,x) &= \sum_{m=1}^{M} \left( f_A(t) g_{C_m}(x) \right) + \sum_{n=1}^{N} \left( f_A^{\theta_n}(t) g_{D_n}(x) \right) \\
&= f_A(t) \sum_{m=1}^{M} \left( g_{C_m}(x) \right) + f_A(t) \sum_{n=1}^{N} \left( \cos\theta_n \ g_{D_n}(x) \right) + \hat{f}_A(t) \sum_{n=1}^{N} \left( \sin\theta_n \ g_{D_n}(x) \right) \\
&= f_A(t) \underbrace{\left( \sum_{m=1}^{M} g_{C_m}(x) + \sum_{n=1}^{N} \left( \cos\theta_n \ g_{D_n}(x) \right) \right)}_{g_A(x)} + \hat{f}_A(t) \underbrace{\sum_{n=1}^{N} \left( \sin\theta_n \ g_{D_n}(x) \right)}_{g_B(x)} \\
&= f_A(t) g_A(x) + \hat{f}_A(t) g_B(x) \\
&= h^{TS}(t,x)
\end{aligned}
\qquad (41)
$$

The physiological interpretation of Eq. 40 is that the cortical cell may receive inputs from many thalamic inputs, not just two, and those many cells need not be identical to each other. Nor do we require that the spectral response fields be related in any way, though they may (they will likely have similar best frequencies due to the tonotopic organization of the auditory system, but they may be of different spectral response field shapes). The inputs have been broken into two groups corresponding to the two terms of Eq. 37. The first input group consists of $M$ ordinary unlagged inputs. The second input group consists of $N$ lagged inputs. The phase lags may all be the same or they may be different.

To restate, we conclude that models which attempt a decomposition of the form

$$h^{TS}(t,x) = \sum_{m=1}^{M}\left(f_A(t)g_{C_m}(x)\right) + \sum_{n=1}^{N}\left(f_{B_n}(t)g_{D_n}(x)\right) \tag{42}$$

where $f_{B_n}(t) \neq f_A^{\theta_n}(t)$ for some $\theta_n$ are ruled out. Such models implicitly violate the temporal symmetry property.

## Multiple Fast Input & Slow Output Model

There is another physiologically motivated relaxation we can allow in our progressive decomposition, to allow the thalamic temporal inputs to not be so carefully matched, as long as the differences do not enter into the cortical computation, e.g. at high frequencies. For instance, we note that the temporal processing of the cell possessing this STRF is well characterized as the convolution of its faster pre-somatic (multiple input) stage and its slower somatic (aggregate processing) stage. We can thus separate the entire temporal component of the cortical processing, $f(t)$, which is a temporal impulse response, into the convolution of impulse responses from a pre-somatic stage, $k(t)$, and somatic stage, $k_A(t)$.

$$f(t) = k(t) * k_A(t) \tag{43}$$

The pre-somatic stage may contain high frequencies (e.g. the output of thalamic ventral MGB neurons) that will not be passed though the slower soma. In filter terminology, the somatic stage is a much lower frequency band-pass filter than the pre-somatic stage (the soma acts as an integrator, which is a form of low-pass filtering). The convolution operation is used since the filters (characterized by impulse responses) are in series. The reason for making this separation is that we can allow each input component $f_i(t)$ to be different as long as the difference is zero when passed through the somatic component (Yeshurun, Wollberg, Dyn, & Allon, 1985),

$$\left(k_i(t) - k_j(t)\right) * k_A(t) = 0, \quad \text{for all } i,j \tag{44}$$

$$k_i(t) * k_A(t) = k_j(t) * k_A(t) = f_A(t), \quad \text{for all } i,j \tag{45}$$

and still obtain a temporally symmetric STRF. Physiologically, we are allowing the thalamic inputs to have greatly differing temporal impulse responses, so long as all the differences appear in timescales *significantly faster* than the somatic processing timescale (Creutzfeldt, Hellweg, & Schreiner, 1980; Rouiller, de Ribaupierre, Toros-Morel, & de Ribaupierre, 1981).

$$h^{TS}(t,x) = \left(\sum_{m=1}^{M}\left(k_{A_m}(t)g_{C_m}(x)\right) + \sum_{n=1}^{N}\left(k_{A_n}^{\theta_n}(t)g_{D_n}(x)\right)\right) * k_A(t) \tag{46}$$

Here the decomposition is broken up into pre-somatic input stage and a somatic processing stage, shown schematically in Figure 10d. The somatic processing stage is the final convolution with the somatic (low-pass) temporal impulse response $k_A(t)$. The input stage looks very similar to the previous decomposition, except that now all the individual input components may have temporal responses that are different from each other, as long as the differences are all for high frequencies (fast timescales), and they remain identical for low frequencies (slow timescales). This is equivalent to the previous decomposition or to that in Eq. 36:

$$h(t,x) = \left( \sum_{m=1}^{M} \left( k_{A_m}(t) g_{C_m}(x) \right) + \sum_{n=1}^{N} \left( k_{A_n}^{\theta_n}(t) g_{D_n}(x) \right) \right) * k_A(t)$$

$$= \sum_{m=1}^{M} \left( \left( k_{A_m}(t) * k_A(t) \right) g_{C_m}(x) \right) + \sum_{n=1}^{N} \left( \left( k_{A_n}^{\theta_n}(t) * k_A(t) \right) g_{D_n}(x) \right)$$

$$= \sum_{m=1}^{M} \left( f_A(t) g_{C_m}(x) \right) + \sum_{n=1}^{N} \left( f_A^{\theta_n}(t) g_{D_n}(x) \right) \tag{47}$$

$$= h^{TS}(t,x)$$

Note the Hilbert rotation of the input stage $k(t)$ is conferred on the entire temporal filter $f(t)$, since

$$k^{\theta}(t) * k_S(t) = \left( \cos\theta \, k(t) + \sin\theta \, \hat{k}(t) \right) * k_A(t)$$

$$= \cos\theta \left( k(t) * k_A(t) \right) + \sin\theta \left( \hat{k}(t) * k_A(t) \right)$$

$$= \cos\theta \, f_A(t) + \sin\theta \, \hat{f}_A(t) \tag{48}$$

$$= f_A^{\theta}(t)$$

which uses the Hilbert transfer property

$$\hat{k}_1(t) * k_2(t) = \mathcal{H}\left[ k_1(t) \right] * k_2(t)$$

$$= \mathcal{H}\left[ k_1(t) * k_2(t) \right] \tag{49}$$

$$= \hat{k}_{12}(t)$$

where $k_1(t) * k_2(t) = k_{12}(t)$. That is, once lagging is performed in the thalamus, the effects of lagging continue up the pathway.

The physiological interpretation of Eq. 46 is that the thalamic inputs are quite unrestricted in their similarity. We do not require that all thalamic inputs have the same temporal impulse response, plus Hilbert rotations of that same temporal impulse response, as would be implied by Eq. 40. We require that only the low frequency components of the thalamic impulse response be the same, and that there be partial lagging in some subset of the population. High frequency components of the temporal impulse responses and all aspects of the spectral response fields are completely unrestricted. In fact, high frequency components of the temporal impulse responses need not even be lagged, since they are filtered out.

To restate, models which attempt a decomposition of the form

$$h^{TS}(t,x) = \left( \sum_{m=1}^{M} \left( k_{A_m}(t) g_{C_m}(x) \right) + \sum_{n=1}^{N} \left( k_{A_n}^{\theta_n}(t) g_{D_n}(x) \right) \right) * k_A(t) \tag{50}$$

where $k_i(t) * k_A(t) \neq k_j(t) * k_A(t)$ for all $i, j$ are ruled out.

From the signal processing point of view, because all the thalamic inputs $k_i(t)$ and $k_j(t)$, as impulse responses, are so fast compared to the slow somatic filter $k_A(t)$ they are convolved with, they are indistinguishable from a pure impulse $\delta(t)$. More rigorously, in the frequency domain, the thalamic input transfer functions (Fourier transform representations of the thalamic input impulse responses) change only gradually over the typical range of the somatic frequencies (a few tens of Hz). A transfer function with approximately constant amplitude and constant phase is approximately the impulse function $\delta(t)$. That the thalamic inputs may also contain delays does not change the result: A transfer function with approximately constant amplitude and linear phase is approximately the delayed impulse function $\delta(t - t_0)$. Even differential delays among inputs can be tolerated if the differences are not too great, since a relative delay shift of 2 ms changes phase by only 4% of a cycle over 20 Hz and less for slower frequencies. For instance, differential delays of a few milliseconds that arise from different thalamic inputs or from different dendritic branches of a cortical neuron are too small to break the temporal symmetry.

Let us summarize the implications of temporal symmetry for thalamic inputs to AI: the thalamic inputs are almost unrestricted, especially in their spectral support, but they must have the same low frequency temporal structure (e.g. approximately constant amplitude and phase linearity for a few tens of Hz). Any lagged inputs (significantly different from zero but not necessarily near 90°), will cause the cortical STRF to lose full separability but maintain temporal symmetry. This is a prediction of the reasonably simple but still physiologically plausible models presented above. There are no constraints imposed by temporal symmetry on the spectral response fields of the thalamic inputs (despite tonotopy in both ventral MGB and AI), nor are there any constraints on the thalamic input impulse response's high frequency (fast) components individually. The main assumptions of the models are that the thalamic inputs are fully separable and that the soma in AI includes a low-pass filter.

## Implications & Interpretation of AI STRF Temporal Properties for Neural Connectivity and Intracortical Inputs to AI

Neurons in AI receive their predominant thalamic (ventral MGB) input in layers III and IV (P. H. Smith & Populin, 2001). Other neurons in a given local circuit receive primarily intracortical inputs, and neurons receiving thalamic inputs receive intracortical inputs as well. Temporal symmetry of the STRF places strong constraints on the nature of these intracortical inputs as well. Unlike the thalamic inputs which area assumed above to be fully separable, we assume that the STRFs of cortical inputs are not fully separable but do have temporal symmetry. In this section, we explore the constraints on the temporal and spectral components of the intracortical inputs to an AI neuron.

## Adding Cortical Feedforward Connections

The first generalization is that we allow the temporally symmetric AI neuron to project to another AI neuron, which itself receives no other input, shown schematically in Figure 11a. This second neuron is then also temporally symmetric, and its STRF is given by the original STRF its input, convolved with the intrinsic impulse response of the second neuron.

$$h^{TS}(t,x) = \left( \sum_{m=1}^{M} \left( k_{A_m}(t) g_{C_m}(x) \right) + \sum_{n=1}^{N} \left( k_{A_n}^{\theta_n}(t) g_{D_n}(x) \right) \right) * k_A(t) * k_2(t) \qquad (51)$$

[Figure 11 about here.]

Before we show the next generalization that maintains temporal symmetry, it is instructive to see a model generalization that does not maintain temporal symmetry, and so is disallowed by the physiology. The sum of two temporally symmetric STRFs is not, in general, temporally symmetric:

$$\begin{aligned}
h_1^{TS}(t,x) &= f_{A1}(t) g_{A1}(x) + \hat{f}_{A1}(t) g_{B1}(x) \\
h_2^{TS}(t,x) &= f_{A2}(t) g_{A2}(x) + \hat{f}_{A2}(t) g_{B2}(x) \\
h(t,x) &= h_1^{TS}(t,x) + h_2^{TS}(t,x) \\
&= f_{A1}(t) g_{A1}(x) + \hat{f}_{A1}(t) g_{B1}(x) + f_{A2}(t) g_{A2}(x) + \hat{f}_{A2}(t) g_{B2}(x) \\
&\neq h^{TS}(t,x)
\end{aligned} \qquad (52)$$

since it generically gives an STRF of rank 4, which cannot be temporally symmetric since then it must be rank 2. There do exist configurations in which there is a special relationship between the several spectral functions, which does result in temporal symmetry, but they are highly restrictive: e.g. when $g_{A1}(x) = g_{A2}(x)$ and $g_{B1}(x) = g_{B2}(x)$. An example of an STRF that is of the form of Eq. 52, where there is no special relationship between the spectral and temporal components of its constituents, is shown Figure 6c. It is not temporally symmetric and is therefore ruled out as a model.

## Adding Cortical Feedback Connections

The next generalization is at a higher level, in that it is the first generalization we consider that has feedback.

$$h_1^{TS}(t,x) = \left( \sum_{m=1}^{M} \left( k_{A_m}(t) g_{C_m}(x) \right) + \sum_{n=1}^{N} \left( k_{A_n}^{\theta_n}(t) g_{D_n}(x) \right) \right) * k_A(t) + h_2^{TS}(t,x)$$
$$h_2^{TS}(t,x) = h_1^{TS}(t,x) * k_2(t)$$

(53)

This is an important generalization, shown schematically in Figure 11b. The proof that the result is temporally symmetric is by causal induction: line two of Eq. 53 guarantees that the STRF of neuron 2 is temporally symmetric with the same spectral functions $g_A(x)$ and $g_B(x)$ of neuron 1. Then since the result of line 1 is the sum of two temporally symmetric STRFs sharing the same $g_A(x)$ and $g_B(x)$, the resulting STRF is also temporally symmetric (and shares the same $g_A(x)$ and $g_B(x)$). A more rigorous proof follows straightforwardly when using the $\text{MTF}_{\text{ST}}$ instead of the STRF.

This illustrates the base of a large class of generalizations: Any network of cortical neurons in one self-contained circuit, including feedforward and feedback connections, will remain temporally symmetric as long as all spectral information arises from a single source that receives all thalamic input. That is, if input into the cortical neural circuit arises, ultimately, from a single source receiving all thalamic input (e.g. one neuron in layer III/IV), then all neurons in that circuit will be temporally symmetric and all will share the same spectral functions $g_A(x)$ and $g_B(x)$. A substantially more complex example is illustrated schematically in Figure 11c.

The converse to this rule of a single source of thalamic inputs to a cortical circuit is that almost any other inputs will break temporal symmetry and are therefore disallowed by the physiology. As shown in Eq. 52, any neuron, if it receives input from two sources whose temporal symmetries are incompatible, will not be temporally symmetric, and is therefore disallowed. Essentially, each local circuit must get its input from a single layer III/IV cell. This is a strong claim: it demands that neurons in the local circuit but not in layer III/IV must also be temporally symmetric. Generally, inputs from neighboring or distant circuits will have a different spectral balance of thalamic inputs and would break temporal symmetry if strongly coupled, at least at time scales important to these STRFs (a few to a few tens of Hz). Only if the neighboring cortical cell were to have a compatible temporal symmetry would it be allowed take part in the circuit without violating temporal symmetry.

Fully separable cells in AI are also well explained by all the above models, except *without the second lagged thalamic population*. All of the results, generalizations, and restrictions apply equally strongly, since anything which breaks temporal symmetry increases the rank, and so does not allow the STRF to be of rank 1 either; and maintaining temporal symmetry is consistent with full separability if there are no lagged inputs to increase the STRF rank from 1 to 2.

## *DISCUSSION*

Temporal symmetry is a property that STRFs may possess, but there is not any mathematical reason that they must. Nevertheless, in the awake case, 85% of neurons measured in AI were temporally symmetric, i.e. $\eta_t > 0.65$, and 96% in the anesthetized population (including those which are fully separable). This property merits serious attention.

Simple models of functional thalamic input to AI, where the temporal functions are unconstrained, or even if identical but with relative delays of more than a few milliseconds, do not possess the property of temporal symmetry, and so must be ruled out. On the other hand, it is quite possible for a model to achieve temporal symmetry: The only restriction on thalamic inputs is that their low-pass filtered versions (at rates below 20 Hz, i.e., those commensurate with cortical integration time constants) must be identical. Additionally, some of the ventral MGB inputs must be (at least partially) lagged in order for the cortical neuron to be temporally symmetric without being fully separable.

Additional constraints are forced upon neurons in AI receiving input from other neurons in AI, whether from simple feedforward or as part of a complicated feedback/feedforward network. Simply taking two temporally symmetric neurons as input will not generate a temporally symmetric neuron. Since the temporal profiles of cortical neurons are not identical, the only way to maintain temporal symmetry is to keep the same spectral balance of the lagged and non-lagged components ($g_A(x)$ and $g_B(x)$ in Eq. 36). The only sure way to accomplish this is to restrict all inputs to the local circuit network to a single input neuron, e.g. in layer III/IV.

Another possibility, not ruled out, is to carefully match the balance of the lagged and non-lagged components even when coming from another column. It seems this would be difficult to match well enough to avoid breaking temporal symmetry, but it is not disallowed by this analysis. It is consistent with the findings of (Read, Winer, & Schreiner, 2001), who note that horizontal connections match not only in best frequency but also in spectral bandwidth.

One can rule out the possibility that temporal symmetry is due a simple interaction between ordinary transmission delays and limited temporal modulation bandwidths. For instance, if the STRF's temporal bandwidth is narrow, an ordinary time delay (with phase linear in frequency) might appear to be constant phase lag. If this effect were important, the temporal symmetry parameter should show a negative correlation with temporal bandwidth. In fact, the correlation between temporal bandwidth (of the first singular value temporal cross-section) and temporal symmetry index is small and positive: 0.24 for the awake case; 0.21 for the anesthetized case.

Of course, the experimental results of Figures 8 and 9 do not show absolute temporal symmetry, only a large degree. Similarly, intracortical connections from neighboring or distant circuits are not ruled out absolutely: rather their strength must be small and their effects more subtle. Indeed, for neurons whose STRF is noisy, the "noise" may be due to inputs from neighboring or distant circuits, whether thalamic or intracortical, which are not time-locked.

In addition, since all the STRFs in this analysis were derived entirely from the steady state portion of the response, this analysis and its models do not apply to neural connections that are only active during onset. This is a limitation of any study of steady state response. The neural connectivity it predicts is a functional connectivity; the connectivity active during the steady state. Nevertheless, the steady state response is important in auditory processing (e.g. for background sounds, auditory streaming, etc.) nor is it is likely that steady state functional connectivity is completely unrelated to other measures of functional connectivity.

It should be noted that the possibility of thalamic inputs being not fully separable but still temporally symmetric (still with all inputs having the same temporal function), is not strictly ruled out mathematically. It amounts to a rearranging of the terms within the largest parentheses of Eq. 46. There is no physiological evidence against the existence of not-fully-separable, temporally symmetric neurons, in auditory thalamus, with all inputs having the same temporal function. In fact, it is possible to shift the models presented here down one level, using input from fully separable neurons in inferior colliculus (IC) instead of thalamus, and using thalamocortical loops instead of intracortical feedback. In this case, it would be necessary to find a mechanism that lags (Hilbert rotates) some of the IC inputs. The essential predictions are the same: all input to the circuit must be through a single source, and any exceptions would break the temporal symmetry, but the source has been moved down on level. However, it does add another layer of complexity and of strong assumptions to the models, so this appears less likely to be relevant.

If, however, ventral MGB neurons are found with substantially more complicated STRF structures (e.g. quadrant separable but not temporally symmetric, or even inseparable without any symmetries, both of which are more complicated than all of the AI neurons described here), then the models proposed here would not be able to accommodate them, and another explanation will have to be found for the temporal symmetry found in AI.

The specific models proposed by the equations above (e.g. Eqs. 37, 40, 46, 51, 53) are somewhat more general than the precise descriptive text that follows from them. For instance, lagged cortical cells are allowed in place of the lagged thalamic cells: e.g. in the macaque visual system there are both lagged and non-lagged V1 cortical neurons, in addition to the same two populations in visual thalamus (De Valois, Cottaris, Mahon, Elfar, & Wilson, 2000). The models above do not distinguish between those cortical inputs and those thalamic inputs. Similarly, the models do not depend on the mechanism by which input neurons are lagged, whether by thalamic inhibition (Mastronarde, 1987a, 1987b) or synaptic depression (Chance, Nelson, & Abbott, 1998); from Eq. 37, the models do not even depend on the sign or magnitude of the phase lag. Nevertheless, this does not invalidate the models' ease of falsifiability.

The example of a feedback network described by Eq. 53 and depicted in Figure 11b is the simplest of systems that include feedback. The more complex model depicted in Figure 11c is serves as an example of a more complex cortical network with feedback, but the class of feedback systems is so much larger than of feedforward systems that it is beyond the scope of this paper to categorize all of them.

One important feature of these models is their ease of falsifiability. Most interactions will break temporal symmetry. Any functionally connected pair of neurons in AI with temporally symmetric STRFs, but with incompatible impulse responses and spectral response fields, would invalidate the model. Finding direct evidence is also possible, and we propose one experiment as an example: Simultaneous recording from two functionally connected neurons in AI with strong temporally symmetric STRFs. While presenting steady state auditory stimuli, electrically stimulate the neuron whose output is the other's input, in a way that is incompatible with temporal symmetry, but compatible with the steady state stimulus. If these models are correct, the STRF of the other neuron will remain well formed but lose temporally symmetry.

Additional caveats are in order, delineating the types of functional connections permitted and disallowed by the measured temporal symmetry. The STRF captures many features of the response in AI, but not all. For instance, typical STRFs in AI only go up to a few tens of Hz (Depireux, Simon, Klein, & Shamma, 2001) since the neurons cannot lock to rates higher than that, yet some spikes within the response show remarkably high temporal precision across different presentations—as fast as a millisecond (see, e.g. Elhilali, Fritz, Klein, Simon, & Shamma, 2004). Since STRFs do not capture effects at this time scale, predictions made from the temporal symmetry of the STRFs cannot place any constraints on functional connections that take effect only at this time scale. Similarly, since the STRFs are measured only from the sustained response and do not use the onset response, constraints cannot be placed on functional connections that occur *only* in the first few hundred milliseconds.

These results have been obtained from STRFs measured by a single method (using TORC stimuli) and still need to be verified for STRFs measured using different stimuli. Use of the SVD truncation (Eq. 21) and calculation of the temporal symmetry index (Eq. 22) is straightforward for STRF from any source. The main caveat in applying this technique to other data is that the STRFs should be sufficiently error free, especially in systematic error (Klein, Simon, Depireux, & Shamma, 2006). Too much error in the STRF estimate leads to the second term in the singular value truncation being dominated by noise, which in turn corrupts the estimate of the temporal symmetry index.

How much of this work generalizes beyond AI? In vision, it appears that temporal and spectral symmetry measurements for spatio-temporal response fields are rarely, if ever, made (these measurement would require the entire spatio-temporal response field to be measured, not just the temporal response at the best spatial frequency, or the converse). It is certainly true that many spatio-temporal response field show strong directional selectivity, and it is clear from Eq. 32, that an STRF which is quadrant separable and purely directional selective is also temporally symmetric.

The population distributions of the temporal symmetry index, shown in Figure 8a, exhibit some differences between the awake and anesthetized populations, the most pronounced of which is the low symmetry tail in the awake population. The similarities are more profound than the differences, however: the large majority of STRFs show pronounced temporal symmetry. This is especially striking when compared with the spectral case, shown in Figure 8b, which has broad distribution in both the awake and anesthetized populations.

Let us also restate an earlier point regarding the validity of drawing conclusions from a linear systems analysis of a system known to have strong non-linearities. Linear systems are a subcategory of those non-linear systems well described by Volterra and Wiener expansions (Eggermont, 1993; Rugh, 1981). If, in particular, the system is well described by a linear kernel followed by a static nonlinearity, the measured STRF is still proportional to that linear kernel (Bussgang, 1952; Nykamp & Ringach, 2002), since the stimuli used here, though discretized in rate and spectral density, are approximately spherical. If the system's nonlinearities are more general, or the static non-linearity is sensitive to the residual non-sphericity of the stimuli, then the measured STRF will contain additional, stimulus-dependent, non-linear contributions. Crucially, it has been shown that those potential higher order non-linearities are not dominant (i.e. the linear properties are robust) in ferret AI, since using several stimulus sets with widely differing spectrotemporal properties produce measured STRFs that do not differ substantially from each other (Klein, Simon, Depireux, & Shamma, 2006). This implies that the STRF measured here is dominantly proportional to a linear kernel (followed by some static non-linearity) and that the analysis performed here is valid. To the extent that there is higher order contamination in the STRF, however, the analysis is affected in unknown ways. For example, there are related neural systems for which the dominant response has not been seen to be linear (Machens, Wehr, & Zador, 2004; Sahani & Linden, 2003).

The most immediate other question is whether ventral MGB neurons do have fully separable STRFs, and whether any have lagged outputs. If neither of these hold true, then it becomes very difficult to explain temporal symmetry, which is so easily broken. Another set of questions revolves around the generation of STRFs themselves. What distinguishes AI neurons that have clean STRFs from those that have noisy STRFs, or none at all? Only some (large) fraction of AI neurons have recognizable STRFs, and only some (large) fraction of those recognizable STRFs have a high signal to noise ratio (no specific statistics have been gathered, and a careful analysis is beyond the scope of this paper). What is the cause of this "noise", and how is it related to (presumably) non-phase-locked inputs?

## *Acknowledgements*

S.A.S. was supported by Office of Naval Research Multidisciplinary University Research Initiative, Grant N00014-97-1-0501. D.A.D. was supported by the National Institute on Deafness and Other Communication Disorders, National Institutes of Health, Grant R01 DC005937. We thank Shantanu Ray for technical assistance in electronics and major contributions in customized software design and Sarah Newman and Tamar Vardi for assistance with training animals. We also thank Alan Saul, and three anonymous reviewers from an earlier submission, for their helpful and insightful comments.

## *Appendix*

### Temporal Symmetry Ambiguities

There are ambiguities and symmetries in the decomposition in Eq. 36:

$$h^{TS}(t,x) = f_A(t)g_A(x) + \hat{f}_A(t)g_B(x).$$ (36)

The transformation parameterized by an amplitude and phase $(\lambda, \theta)$

$$f_A(t) \rightarrow \lambda f_A^{\theta}(t) = \lambda\Big(\cos\theta\, f_A(t) + \sin\theta\, \hat{f}_A(t)\Big)$$

$$\hat{f}_A(t) = f_A^{\pi/2}(t) \rightarrow \lambda f_A^{\theta+\pi/2}(t) = \lambda\Big(-\sin\theta\, f_A(t) + \cos\theta\, \hat{f}_A(t)\Big)$$

$$g_A(x) \rightarrow \lambda^{-1}\big(\cos\theta\, g_A(x) + \sin\theta\, g_B(x)\big)$$ (A1)

$$g_B(x) \rightarrow \lambda^{-1}\big(-\sin\theta\, g_A(x) + \cos\theta\, g_B(x)\big)$$

leaves Eq. 36 invariant. This indicates a two-fold symmetry, which here means that this decomposition is ambiguous (non-unique). The amplitude symmetry, parameterized by $\lambda$, demonstrates that the power (strength) of an STRF cannot be ascribed to either the spectral or the temporal cross section. A standard method of removing this ambiguity is to normalize the cross sections and include an explicit amplitude:

$$h^{TS}(t,x) = \Lambda_A v_A(t) u_A(x) + \Lambda_B \hat{v}_A(t) u_B(x) \tag{A2}$$

where the $u_i$ and $v_i$ are of unit norm. This method is used in SVD.

The rotation transformation parameter $\theta$ merely establishes which of the possible rotations of $f_A^\theta(t)$ we choose to compare against all others. For example, in Figure 7, the temporal cross-section at the best frequency (i.e. the frequency at which the STRF reaches its most extreme value) is chosen to be the standard impulse response against which all others are compared. To do this, we choose $\theta$ such that $h^{TS}(t,x_{BF}) = f_A(t) g_A(x_{BF})$ and $\lambda$ such that $g_A(x_{BF}) = 1$.

## Quadrant Separable Spectro-temporal Modulation Transfer Functions

For mathematical convenience, we name and define the cross sectional functions referred to without names above in the section *Directionality and the Quadrant Separability*. In quadrant 1, where both $w$ and $\Omega$ are positive, $H(w,\Omega) = F_1^+(w) G_1^+(\Omega)$, where the index 1 refers to quadrant 1 and the + indicates that the function is defined only for positive argument. In quadrant 2, where $w$ is negative and $\Omega$ is positive, $H(w,\Omega) = F_2^{+*}(-w) G_2^+(\Omega)$, where the index 2 refers to quadrant 2, the + indicates that the function is defined only for positive argument, $-w$ is positive, and the complex conjugation makes later equations significantly simpler. Thus, a quadrant separable MTF$_{ST}$ can be expressed as (valid in all 4 quadrants)

$$H^{QS}(w,\Omega) = F_1^+(w)\Theta(w)G_1^+(\Omega)\Theta(\Omega)$$
$$+ F_2^+(w)\Theta(w)G_2^{+*}(-\Omega)\Theta(-\Omega)$$
$$+ F_1^{+*}(-w)\Theta(-w)G_1^{+*}(-\Omega)\Theta(-\Omega)$$
$$+ F_2^{+*}(-w)\Theta(-w)G_2^{+*}(\Omega)\Theta(\Omega)$$

(A3)

where $\Theta()$ is the step function (1 for positive argument and 0 for negative argument). In Eq. A3 it is explicit that the spectro-temporal modulation transfer function in quadrants 3 $(w<0, \Omega<0)$ and 4 $(w>0, \Omega<0)$ is complex conjugate to quadrants 1 and 2 respectively.

This can be written more compactly as,

$$H^{QS}(w,\Omega) = F_1(w)G_1(\Omega) + F_2(w)G_2(\Omega) - \hat{F}_1(w)\hat{G}_1(\Omega) + \hat{F}_2(w)\hat{G}_2(\Omega)$$

(A4)

where

$$F_i(w) = F_i^+(w)\Theta(w) + F_i^{+*}(-w)\Theta(-w)$$
$$G_i(\Omega) = G_i^+(\Omega)\Theta(\Omega) + G_i^{+*}(-\Omega)\Theta(-\Omega)$$

(A5)

are defined for both positive and negative argument (and are complex conjugate symmetric by construction), $i = (1,2)$, and the Hilbert transform is defined in transfer function space as

$$\hat{F}(w) = j \, \mathrm{sgn}(w) \, F(w)$$
$$\hat{G}(\Omega) = j \, \mathrm{sgn}(\Omega) \, F(\Omega)$$

.

(A6)

Eq. A4 demonstrates that a quadrant separable spectro-temporal modulation transfer function can be decomposed into the sum of 4 fully separable spectro-temporal modulation transfer functions where the last two terms are fully determined by the first two and the requirement of quadrant separability. This is the first time we see that a generic quadrant separable MTF$_{ST}$ (and its STRF) is rank 4, and not 2. Roughly speaking it comes from the combination of two independent product terms in quadrants 1 and 2, *plus* the restriction that the MTF$_{ST}$ must be complex conjugate symmetric.

A equivalent decomposition equally concise, but more amenable to comparison with definitions motivated earlier, of a quadrant separable spectro-temporal modulation transfer function into the sum of 4 fully separable spectro-temporal modulation transfer functions is

$$H^{QS}(w,\Omega) = F_A(w)G_A(\Omega) + \hat{F}_B(w)\hat{G}_B(\Omega) + \hat{F}_A(w)G_B(\Omega) + F_B(w)\hat{G}_A(\Omega) \tag{A7}$$

where

$$
\begin{aligned}
F_A(w) &= \tfrac{1}{\sqrt{2}}\big(F_1(w) + F_2(w)\big) \\
F_B(w) &= \tfrac{1}{\sqrt{2}}\big(-\hat{F}_1(w) + \hat{F}_2(w)\big) \\
G_A(\Omega) &= \tfrac{1}{\sqrt{2}}\big(G_1(\Omega) + G_2(\Omega)\big) \\
G_B(\Omega) &= \tfrac{1}{\sqrt{2}}\big(-\hat{G}_1(\Omega) + \hat{G}_2(\Omega)\big).
\end{aligned}
\tag{A8}
$$

Taking the inverse Fourier Transform gives the canonical form of the quadrant separable STRF

$$H^{QS}(t,x) = f_A(t)g_A(x) + \hat{f}_B(t)\hat{g}_B(x) + \hat{f}_A(t)g_B(x) + f_B(t)\hat{g}_A(x) \tag{A9}$$

which is also Eq. 29.

## Quadrant Separability Ambiguities

There are ambiguities and symmetries in the decomposition in Eq. A3 (and correspondingly Eq. A4 and A7). The transformation parameterized by a pair of amplitudes and phases $(\Lambda_1, \theta_1, \Lambda_2, \theta_2)$

$$F_1^+ \rightarrow F_1^+ \Lambda_1 \exp(j\theta_1)$$
$$G_1^+ \rightarrow G_1^+ \Lambda_1^{-1} \exp(-j\theta_1)$$
$$F_2^+ \rightarrow F_2^+ \Lambda_2 \exp(j\theta_2) \qquad \text{(A10)}$$
$$G_2^+ \rightarrow G_2^+ \Lambda_2^{-1} \exp(-j\theta_2)$$

or equivalently,

$$F_1(w) \rightarrow \Lambda_1 F_1^{\theta_1}(w)$$
$$G_1(\Omega) \rightarrow \Lambda_1^{-1} G_1^{-\theta_1}(\Omega)$$
$$F_2(w) \rightarrow \Lambda_2 F_2^{\theta_2}(w) \qquad \text{(A11)}$$
$$G_2(\Omega) \rightarrow \Lambda_2^{-1} G_2^{-\theta_2}(\Omega)$$

leaves Eq. A3 invariant, representing an ambiguous/non-unique decomposition. The spectro-temporal version of the transformation is

$$f_1(t) \rightarrow \Lambda_1 f_1^{\theta_1}(t)$$
$$g_1(x) \rightarrow \Lambda_1^{-1} g_1^{-\theta_1}(x)$$
$$f_2(t) \rightarrow \Lambda_2 f_2^{\theta_2}(t) \qquad \text{(A12)}$$
$$g_2(x) \rightarrow \Lambda_2^{-1} g_2^{-\theta_2}(x)$$

The amplitude symmetry, parameterized by $\Lambda_1$ and $\Lambda_2$, demonstrates that the amplitude (or power) in one quadrant of a spectro-temporal modulation transfer function cannot be ascribed to either the spectral or the temporal cross section. Again, a standard method of removing this ambiguity is to normalize the cross sections and include an explicit amplitude:

$$H_{ST}^{QS}(w,\Omega) = V_1(w)\Lambda_1 U_1(\Omega) + V_2(w)\Lambda_2 U_2(\Omega) - \hat{V}_1(w)\Lambda_1\hat{U}_1(\Omega) + \hat{V}_2(w)\Lambda_2\hat{U}_2(\Omega) \quad \text{(A13)}$$

where the $U_i$ and $V_i$ are of unit norm. This method is used in Singular Value Decomposition (SVD). The spectro-temporal version is:

$$H^{QS}(t,x) = v_1(t)\Lambda_1 u_1(x) + v_2(t)\Lambda_2 u_2(x) - \hat{v}_1(t)\Lambda_1 \hat{u}_1(x) + \hat{v}_2(t)\Lambda_2 \hat{u}_2(x) \, . \qquad \text{(A14)}$$

The phase ambiguity, parameterized by $\theta_1$ and $\theta_2$, can be useful. By judicious choice of this transformation, it can always be arranged that $F_2$ be orthogonal to $\hat{F}_1$ and $G_2$ be orthogonal to $\hat{G}_1$ simultaneously. This is useful when performing detailed calculations (e.g. the analytic expression below for the singular values of a quadrant separable spectro-temporal modulation transfer function) because it explicitly decomposes the spectro-temporal modulation transfer function, viewed as a linear operator, into components that act on orthogonal subspaces.

## Rank of Quadrant Separable Spectro-temporal Response Fields

This section uses SVD to prove that the rank of a generic quadrant separable STRF is 4.

We begin with Eq. A14. In this form, the symmetry generated by $\Lambda_1$ and $\Lambda_2$ in Eq. A11 has already been fixed but we can still use the symmetry generated by $\theta_1$ and $\theta_2$ in Eq. A11 to simplify later calculations. We quantify the non-orthogonality of the terms in Eq. A14 by the 4 angles.

$$\begin{aligned}
\cos\theta_V &= v_1 \cdot v_2 = \hat{v}_1 \cdot \hat{v}_2 \\
\cos\theta_U &= u_1 \cdot u_2 = \hat{u}_1 \cdot \hat{u}_2 \\
\cos\varphi_V &= v_1 \cdot \hat{v}_2 = -\hat{v}_1 \cdot v_2 \\
\cos\varphi_U &= u_1 \cdot \hat{u}_2 = -\hat{u}_1 \cdot u_2
\end{aligned} \qquad \text{(A15)}$$

where $w_i \cdot w_j := \int w_i w_j$, and $v_i \cdot v_j = \hat{v}_i \cdot \hat{v}_j = \delta_{ij} = u_i \cdot u_j = \hat{u}_i \cdot \hat{u}_j$. By making an explicit transformation with $\theta_1$ and $\theta_2$ from Eq. A12 such that

$$\tan(\theta_1 - \theta_2) = -\frac{\cos\varphi_V}{\cos\theta_V}$$

$$\tan(\theta_1 + \theta_2) = \frac{\cos\varphi_U}{\cos\theta_U}$$

(A16)

then it can be shown that *after* the transformation

$$v_1 \cdot \hat{v}_2 = 0 = u_1 \cdot \hat{u}_2 \,,$$

(A17)

and there are new values for $\cos\theta_V = v_1 \cdot v_2$ and $\cos\theta_U = u_1 \cdot u_2$ as well. This step is not necessary but it reduces the complexity of the following equations. To avoid double counting, we also need to restrict the range of either $\theta_U$ or $\theta_V$ (just as on a sphere, we let the azimuthal angle run over the entire equator but we enforce the polar angle to only run over half a great circle of longitude). We do this by enforcing that $\cos\theta_U$ be positive.

We then apply SVD to Eq. A14 with the constraint of Eq. A17. The singular values $\lambda_1$, $\lambda_2$, $\lambda_3$, and $\lambda_4$ are the eigenvalues of the operator formed by taking the inner product of the STRF with itself along the $w$ axis.

$$\int dt \, h^{QS}(t,x) \, h^{QS\dagger}(t,x') =$$
$$\Lambda_1^2 u_1(x)u_1(x') + \Lambda_2^2 u_2(x)u_2(x') + \Lambda_1\Lambda_2 \cos\theta_V \big(u_1(x)u_2(x') + u_2(x)u_1(x')\big)$$
$$+ \Lambda_1^2 \hat{u}_1(x)\hat{u}_1(x') + \Lambda_2^2 \hat{u}_2(x)\hat{u}_2(x') - \Lambda_1\Lambda_2 \cos\theta_V \big(\hat{u}_1(x)\hat{u}_2(x') + \hat{u}_2(x)\hat{u}_1(x')\big)$$

(A18)

Each line of the right-hand-side of Eq. A18 can then be orthogonalized separately. In fact, after we orthogonalize the first line, the second line can be orthogonalized by substituting $\cos\theta_V \to -\cos\theta_V$ and taking the Hilbert transform of each component function. To orthogonalize the first line we look for the eigenvalues $\lambda^2$ and coefficients $c_1$ and $c_2$ such that

$$\zeta(x) = c_1 u_1(x) + c_2 u_2(x) \tag{A19}$$

is the eigenfunction solving the eigenvalue equation

$$
\begin{aligned}
\int dx' &\big(\Lambda_1^2 u_1(x) u_1(x') + \Lambda_2^2 u_2(x) u_2(x') \\
&+ \Lambda_1 \Lambda_2 \cos\theta_V \big(u_1(x) u_2(x') + u_2(x) u_1(x')\big)\big)\zeta(x') \\
&= \lambda^2 \zeta(x)
\end{aligned} \tag{A20}
$$

Putting Eq. A19 into Eq. A20 and expanding, gives a pair of solutions defined (up to normalization) by the two roots of the quadratic equation

$$\left(\frac{c_2}{c_1}\right)^2 \big(\Lambda_1^2 \cos\theta_U + \Lambda_1 \Lambda_2 \cos\theta_V\big) + \left(\frac{c_2}{c_1}\right)\big(\Lambda_1^2 - \Lambda_2^2\big) - \big(\Lambda_2^2 \cos\theta_U + \Lambda_1 \Lambda_2 \cos\theta_V\big), \tag{A21}$$

and the corresponding two solutions of

$$\lambda^2 = \Lambda_1^2 + \Lambda_1 \Lambda_2 \cos\theta_V \cos\theta_U + \left(\frac{c_2}{c_1}\right)\big(\Lambda_1^2 \cos\theta_U + \Lambda_1 \Lambda_2 \cos\theta_V\big). \tag{A22}$$

Solving similarly for the eigenfunctions and eigenvalues of the second line of the right-hand-side of Eq. A12 gives the same as Eq. A21 and A22 but with $\cos\theta_V \rightarrow -\cos\theta_V$.

The eigenfunctions of $w$ are given by using the operator formed by taking the inner product of the STRF with itself along the $\Omega$ axis

$$\int dx\, h^{QS}(t,x)\, h^{QS\dagger}(t',x) \tag{A23}$$

and give the same eigenvalues.

Explicitly, the 4 eigenvalues $\lambda^2$ are given by the 4 solutions of

$$\lambda^2 = \frac{\Lambda_1^2 + \Lambda_2^2}{2} + \Lambda_1 \Lambda_2 \cos\theta_V \cos\theta_U$$
$$\pm \sqrt{\left(\frac{\Lambda_1^2 + \Lambda_2^2}{2}\right)^2 - \Lambda_1^2 \Lambda_2^2 \left(1 - \cos^2\theta_V - \cos^2\theta_U\right) + \left(\Lambda_1^2 + \Lambda_2^2\right)\Lambda_1 \Lambda_2 \cos\theta_V \cos\theta_U}$$
(A24)

and the same with $\cos\theta_V \rightarrow -\cos\theta_V$. The 4 singular values are given by the 4 positive square roots of $\lambda^2$. Because the system is explicitly orthogonal, the rank is given by the number of non-zero values of $\lambda^2$. Generically, the four values of $\lambda^2$ are non-zero, with the only exceptions described below.

# Lower Rank Quadrant Separable Spectro-temporal Response Fields

This section uses SVD to prove that if a quadrant separable STRF has rank less than the generic rank of four, then this is caused by specific symmetries, and the rank must be reduced to 2 (or 1).

Label the 4 eigenvalues $\lambda^2$ as $\lambda_{A\pm}^2$ for the two solutions to (A24) and $\lambda_{B\pm}^2$ for the two solutions to with to Eq. A24 but with $\cos\theta_V \rightarrow -\cos\theta_V$.

First we consider the case that neither $\Lambda_1$ nor $\Lambda_2$ vanish (we will consider the alternative case later). Define $\Lambda_0 = \sqrt{\Lambda_1 \Lambda_2}$ and

$$\alpha^2 = \frac{1}{2}\left(\frac{\Lambda_1^2}{\Lambda_2^2} + \frac{\Lambda_2^2}{\Lambda_1^2}\right)$$
(A25)

where it can be shown that $\alpha^2 > 1$ for any $\Lambda_1$ and $\Lambda_2$. Then the SVD eigenvalues can be written as

$$\lambda^2 = \Lambda_0^2 \begin{pmatrix} \alpha^2 + \cos\theta_V \cos\theta_U \\ \pm\sqrt{\left(\alpha^2 + \cos\theta_V \cos\theta_U\right)^2 - \left(1 + \cos\theta_V \cos\theta_U\right)^2 + \left(\cos\theta_V + \cos\theta_U\right)^2} \end{pmatrix} \quad \text{(A26)}$$

and the same with $\cos\theta_V \rightarrow -\cos\theta_V$. Note for reference below, that, since $\alpha^2 > 1$, the top row expression of Eq. A26 is non-negative, as is the expression under the square root. It can also be shown that the top row expression is greater than or equal to the square root expression, guaranteeing that all of $\lambda_{A\pm}^2$ and $\lambda_{B\pm}^2$ are non-negative.

The loss of rank (from 4 to a smaller number) corresponds to at least one of the eigenvalues vanishing, since the rank is the number of non-zero eigenvalues. Because of the non-negativity properties of the components of Eq. A26, we always have $\lambda_{A+}^2 > \lambda_{A-}^2$ and $\lambda_{B+}^2 > \lambda_{B-}^2$, and so if only one of the eigenvalues vanish it must be $\lambda_{A-}^2$ or $\lambda_{B-}^2$. First consider the case of $\lambda_{A-}^2 = 0$. This can only occur when the top row expression of Eq. A26 cancels the square root expression exactly, or since they are both non-negative, when their squares cancel exactly. This condition, when simplified, can be written

$$\left(1 - \cos^2\theta_V\right)\left(1 - \cos^2\theta_U\right) = 0 \quad \text{(A27)}$$

which can occur only when $\cos^2\theta_V = 1$ or $\cos^2\theta_U = 1$.

The case $\cos^2\theta_V = 1$ results in only 2 non-zero eigenvalues, with values $\lambda^2 = 2\Lambda_0^2\left(\alpha^2 \pm \cos\theta_U\right)$, and 2 zero eigenvalues, $\lambda_{A-}^2 = \lambda_{B-}^2 = 0$, giving an STRF of rank 2 (not rank 3, the guess associated with the initial assumption of $\lambda_{A-}^2 = 0$). From Eq. A15, this also results in $v_2(t) = \pm v_1(t)$, which is necessary and sufficient for temporal symmetry: the STRF has only one temporal function and its Hilbert transform.

Analogously, the case $\cos^2 \theta_U = 1$ results in spectral symmetry. There are only two non-zero eigenvalues, with values $\lambda^2 = 2\Lambda_0^2 \left( \alpha^2 \pm \cos \theta_V \right)$, and two zero eigenvalues, $\lambda_{A-}^2 = \lambda_{B-}^2 = 0$, giving an STRF of rank 2.

Considering the case of $\lambda_{B-}^2 = 0$ leads to the identical conclusions.

Aside from the degenerate case in which only one non-zero eigenvalue remains (which is the fully separable case), the only other possibility of low rank arises from when either $\Lambda_1$ or $\Lambda_2$ vanishes. If $\Lambda_1 = 0$, then the quadrant separable STRF of Eq. A14 is explicitly in the form of Eq. 32, giving this STRF the symmetry of pure directional selectivity. Eq. A24 simplifies dramatically, results in only two non-zero eigenvalues (both with value $\Lambda_1^2$), giving a rank 2 STRF. Similarly, if $\Lambda_2 = 0$, The STRF is purely directionally selective in the opposite direction, resulting in only two non-zero eigenvalues (with value $\Lambda_2^2$), and also rank 2.

Thus we have proven that whenever the rank of a quadrant separable STRF is less than 4, it must be: rank 2 and temporally symmetric; rank 2 and spectrally symmetric; rank 2 and directionally selective; or rank 1 and fully separable.

# *Figure Captions*

**Figure 1**: **a**. A simulated fully separable STRF, with spectral and temporal one-dimensional cross-sections as inserts (all horizontal slices have the same profile, as do vertical slices). **b**. A simulated high rank STRF with no particular symmetries. **c**. A simulated temporally symmetric and quadrant separable STRF of rank 2. This STRF's symmetry is not obviously visible. This STRF was created by adding to the STRF in **a** the same STRF except with the spatial cross-section shifted upward by 3/4 octave and the temporal cross-section Hilbert-rotated (see below) by 30°.

**Figure 2**: **a**. An experimentally measured STRF, with several spectral and temporal one-dimensional cross-sections. **b**. The same STRF interpreted as the spectrogram of an optimal stimulus. **c**. An intricate stimulus, and how different areas of the stimulus spectrogram contribute to the neurons firing rate at any given moment. (Single Unit/Awake: z004b03-p-tor.a1-2)

**Figure 3**: An example of a function and its Hilbert transform. **a**. A function (in black) overlaid with its Hilbert transform (in gray); the two are orthogonal. **b**. The magnitude of the Fourier transform of the function (in black) overlaid with the magnitude of the Fourier transform of its Hilbert transform (in gray); they overlap exactly. **c**. The phase of the Fourier transform of the function (in black) overlaid with the phase of the Fourier transform of its Hilbert transform (in gray); the difference is exactly ±90° (dashed line).

**Figure 4**: **a**. The simulated temporally symmetric and quadrant separable STRF from Figure 1c, and five fixed-frequency cross-sections, corresponding

to five temporal impulse responses. **b**. The same five impulse responses but individually Hilbert-rotated and rescaled. **c**. The same Hilbert-rotated and rescaled impulse responses superimposed. The Hilbert-rotation phases were calculated by taking the negative phase of the complex correlation coefficient between the analytic signal of each temporal cross-section and the analytic signal of the 4th temporal cross-section.

**Figure 5**: **a**. The simulated fully separable $MTF_{ST}$ generated by the STRF in Figure 1a. Phase is given by hue (scale on right), amplitude by intensity. Since the STRF is fully separable, the $MTF_{ST}$ is separable as well, both in amplitude (all horizontal slices have the same intensity profiles, as do vertical slices), and in phase. Separability leads to a phase profile that is the direct sum of a purely temporally dependent phase and a purely spectrally dependent phase. When the phase is primarily linear, as it is for STRFs well localized in spectrum and time, the phase profile is diagonal, with slope determined by the location in spectrum and time of the STRF. **b**. The simulated $MTF_{ST}$ generated by the high rank STRF in Figure 1b. Phase as in a. Since there is no particular symmetry in the STRF, there is no particular symmetry in the $MTF_{ST}$. To the extent that the STRF is localized in spectrum and time, the phase slope is approximately constant. **c**. The simulated $MTF_{ST}$ generated by the temporally symmetric and quadrant separable STRF of rank 2. The symmetry is now more visible than in Figure 1. This $MTF_{ST}$ is somewhat directionally selective: quadrant 2 (characterizing responses to sounds with an upward spectral glide) is strong, and clearly separable within the quadrant; quadrant 1 (characterizing responses to sounds with an downward spectral glide) is weak. Quadrants 3 and 4 are the complex conjugates of quadrants 1 and 2, by Eq. 28.

**Figure 6**: **a**. An STRF equal to the sum of two simulated fully separable STRFs, identical to each other except translated in time and spectrum (the one with lower best frequency and shorter delay is displayed in Figure 1a). This results in a (not fully separable) strongly velocity selective STRF. **b**. The MTF$_{ST}$ of the STRF in a. The spectro-temporal modulation transfer function is clearly not a vertical column-horizontal row product within each quadrant, and therefore the entire spectro-temporal modulation transfer function cannot be quadrant separable. Because the spectro-temporal modulation transfer function is not quadrant separable, it cannot be temporally symmetric. Phase is given by hue (scale on right); amplitude is given by intensity. **c**. An STRF equal to the sum of two simulated temporally symmetric STRFs: identical to each other except translated in time and spectrum (the one with higher best frequency and shorter delay is displayed in Figure 1c). **d**. The first 10 singular values (from SVD) of the STRF show a rank of 4, which cannot be temporally symmetric (temporal symmetry requires a rank of 2).

**Figure 7**: Temporal symmetry demonstrated in the experimentally measured STRFs of 2 example neurons. **a**. A rank 2 (not fully separable) STRF and five fixed-frequency cross-sections, giving five temporal impulse responses. **b**. The same five impulse responses but Hilbert-rotated, and rescaled to have equal power, to that of the impulse response at the best frequency (left panel), and the same Hilbert-rotated and rescaled impulse responses superimposed (right panel). The temporal symmetry index for this neuron is $\eta_t = 0.90$. **c**. Another rank 2 STRF, and five cross-sections (left panel), and the corresponding five impulse responses, Hilbert-rotated, rescaled to have equal power, and superimposed. This temporal symmetry is $\eta_t = 0.73$, illustrating that the temporal symmetry index need not be overly close to unity to demonstrate temporal symmetry. (Single Units/Awake: D2-4-03-p-c.a1-3, R2-6-03-p-2-a.a1-2).

**Figure 8**: The population distributions of the symmetry indices for all neurons with rank 2 (not fully separable) STRFs: awake $N = 70$ (black), and anesthetized $N = 22$ (gray). **a**. The population distributions of temporal symmetry index. The populations are heavily biased toward high temporal symmetry; c.f. the temporal symmetry index of the STRFs in Figure 7 range from 0.73 to 0.90. In the awake population, 51 neurons had temporal symmetry index greater than 0.65; 20 in the anesthetized population. **b**. For comparison, the same statistics but for spectral symmetry. The population is spread over the full range of values, despite potential tonotopic arguments for a narrow distribution near 1.

**Figure 9**: **a**. The population distributions of the correlations for all rank 2 neurons between their rank 2 and quadrant separable estimates: awake (black), and anesthetized (gray). The high correlations indicate that the quadrant separable truncation may be of rank 2, evidence of temporal symmetry. **b**. For comparison, the distributions of the comparable correlations. (Right) The population distributions of the correlations for all neurons between their rank 4 and quadrant separable estimates: awake (black), and anesthetized (gray). The correlations become worse for both populations, despite the fact that generic quadrant separable STRFs are of rank 4, providing more evidence that the quadrant separable STRFs are of rank 2, implying temporal symmetry. (Center) The populations of permuted STRFs: the rank 2 estimate of each STRF is correlated with the quadrant separable estimate of every other STRF: awake (dark gray hash), and anesthetized (light gray hash). The population is scaled by the ratio of the number of STRFs to the number of permuted STRF pairs.

**Figure 10**: Schematics depicting simple models represented by equations below. **a**. Summing the inputs of two fully symmetric (FS) cells whose temporal functions are in quadrature ($\theta = 0$, $\theta = \pi/2$) results in a temporally symmetric (TS) cell (Eq. 36). **b**. Summing the inputs of two fully symmetric (FS) cells whose temporal functions are in *partial* quadrature ($\theta = 0$, $\theta \neq 0$) still results in a temporally symmetric (TS) cell (Eq. 37). **c**. Summing the inputs of *many* fully symmetric (FS) cells whose temporal functions are in partial quadrature still results in a temporally symmetric (TS) cell (Eq. 40). **d**. Summing the inputs of many fully symmetric (FS) cells whose temporal functions are in partial quadrature, with differing impulse responses (at high frequencies), input into a neuron whose somatic impulse response is slow, results in a temporally symmetric (TS) cell (Eq. 46). (Inset) A spectral schematic of the low-pass nature of the slow somatic impulse response: $K_A(f)$ is the Fourier transform of $k_A(t)$.

**Figure 11**: Schematics depicting models that are more complex. **a** Using the output of a temporally symmetric (TS) neuron as sole input to another neuron results in a temporally symmetric (TS) neuron (Eq. 51). **b** Feedback from a such a temporally symmetric neuron whose sole source is the first temporally symmetric neuron is still self-consistently temporally symmetric (Eq. 53). **c** Multiple examples of feedback and feedforward: The initial neuron "TS 1" provides temporal symmetry to all other neurons in the network due to its role as sole input for the network. All other neurons inherit the temporal symmetry, and the feedback is also self-consistently temporally symmetric.

## *Tables and Table Captions*

| | Awake | | Anesthetized | |
|---|---|---|---|---|
| | # | Percent | # | Percent |
| Rank 1 | 72 | 50% | 49 | 67% |
| Rank 2 | 70 | 48% | 22 | 30% |
|    Temporally Symmetric | 51 | | 21 | |
|    Non-Temporally Symmetric | 19 | | 1 | |
| Rank 3 | 3 | 2% | 2 | 3% |
| *Total* | *145* | *100%* | *73* | *100%* |
| Rank 1 + Rank 2 Temporally Symmetric | 123 | 85% | 70 | 96% |
| Rank 2 Non-Temporally Symmetric + Rank 3 | 22 | 15% | 3 | 4% |
| *Total* | *145* | *100%* | *73* | *100%* |

Table 1: The population distributions of STRFs recorded from all neuron, according to animal state (Awake/Anesthetized), STRF rank, and temporal symmetry.

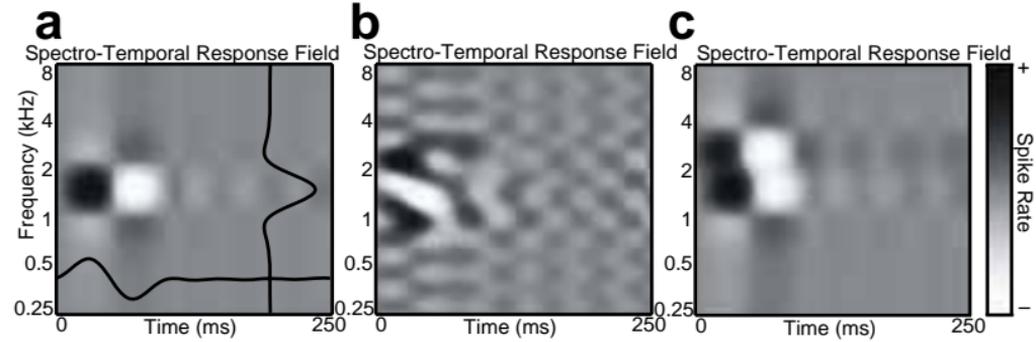

Figure 1

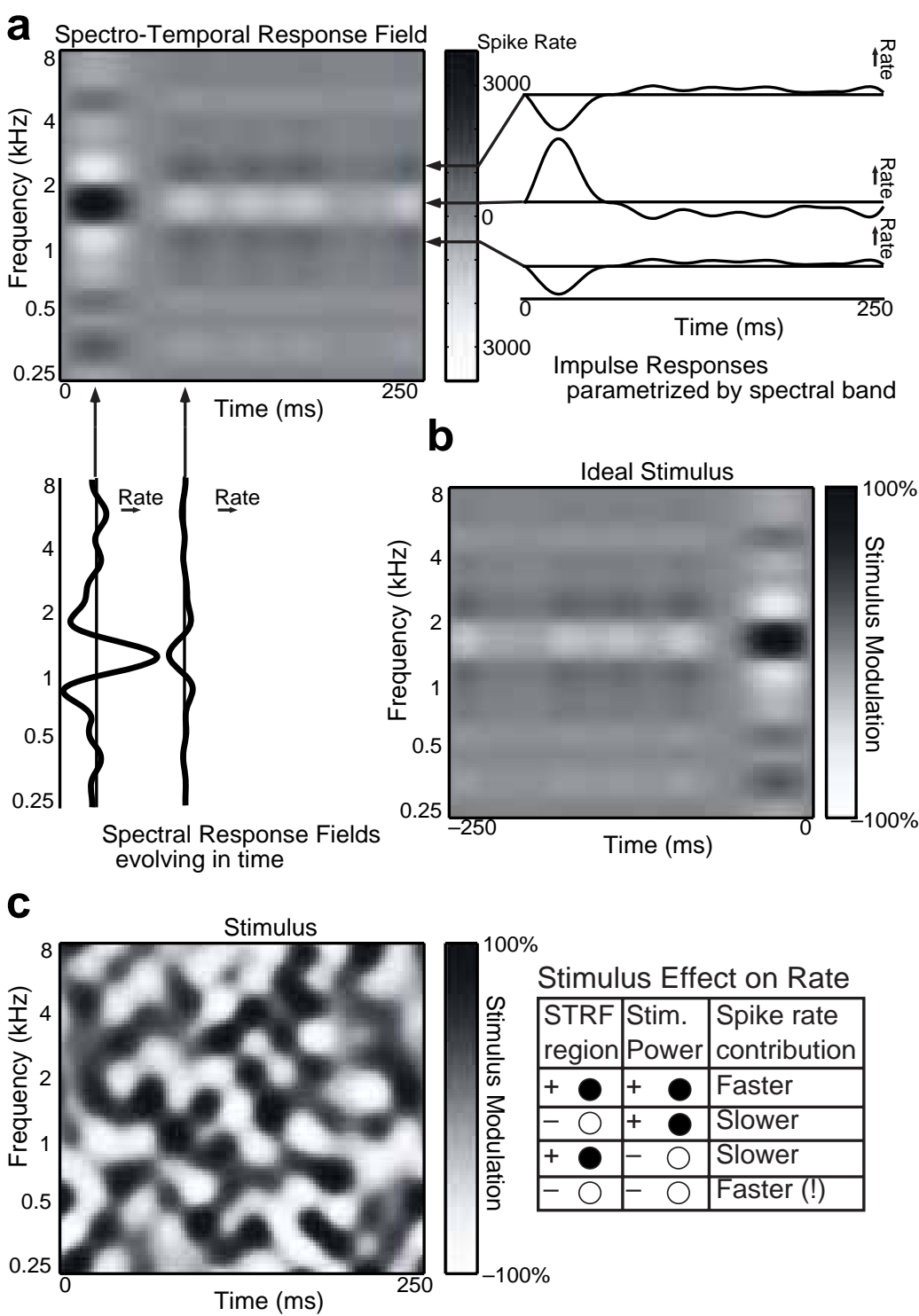

**a** Spectro-Temporal Response Field

Spike Rate

Impulse Responses
parametrized by spectral band

Spectral Response Fields
evolving in time

**b** Ideal Stimulus

**c** Stimulus

Stimulus Effect on Rate

| STRF region | Stim. Power | Spike rate contribution |
|---|---|---|
| + ● | + ● | Faster |
| − ○ | + ● | Slower |
| + ● | − ○ | Slower |
| − ○ | − ○ | Faster (!) |

Figure 2

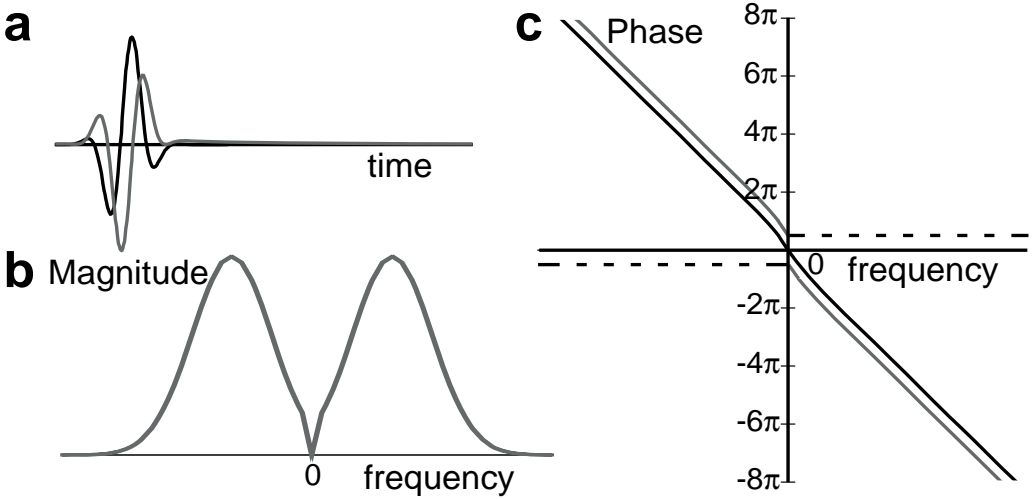

Figure 3

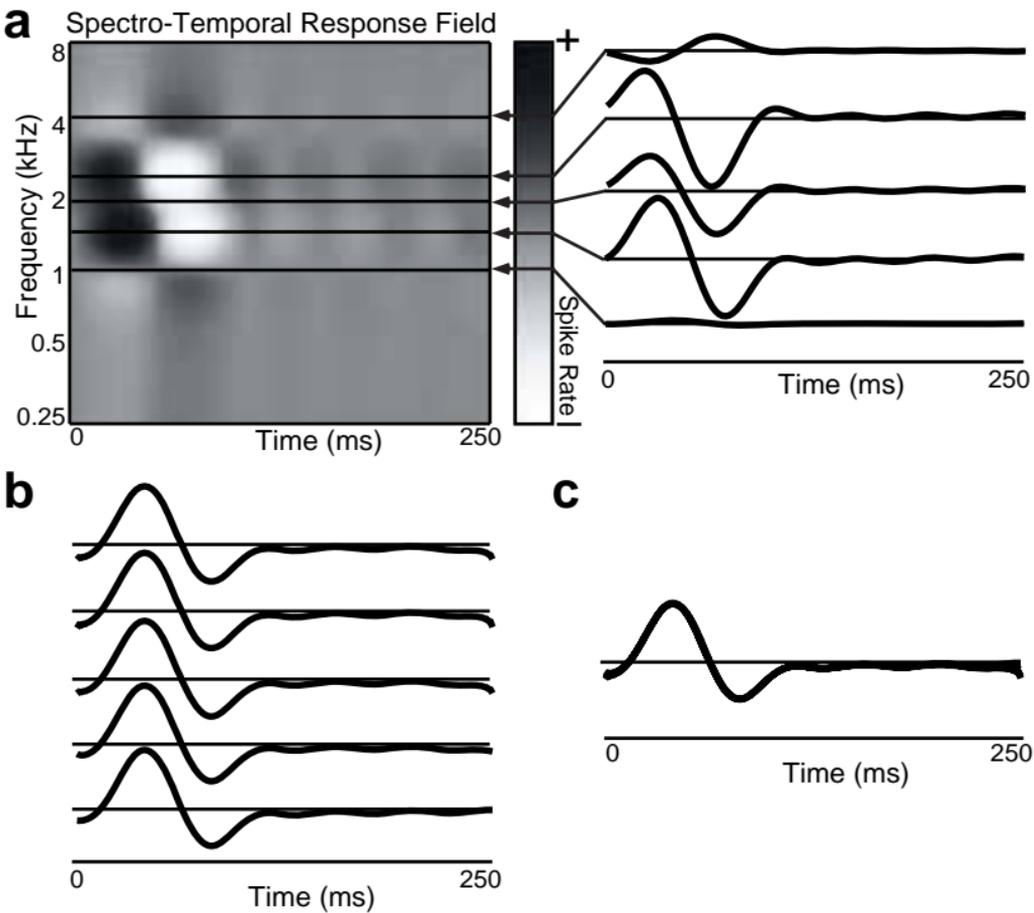

Figure 4

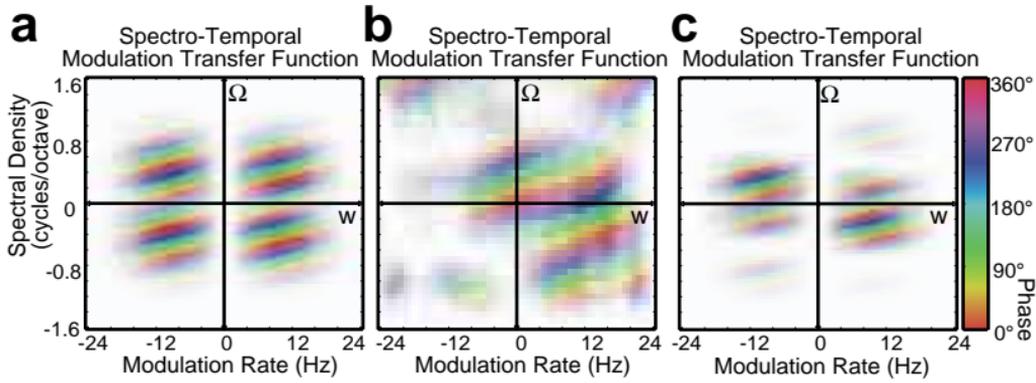

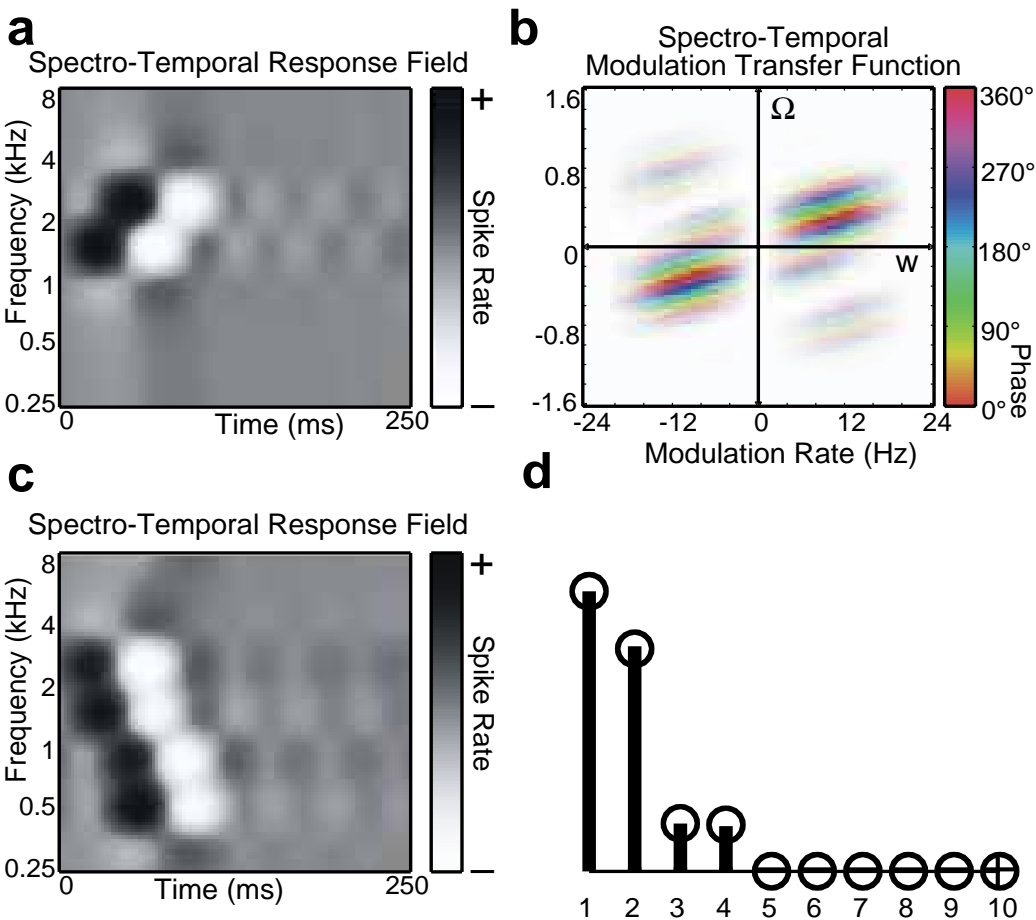

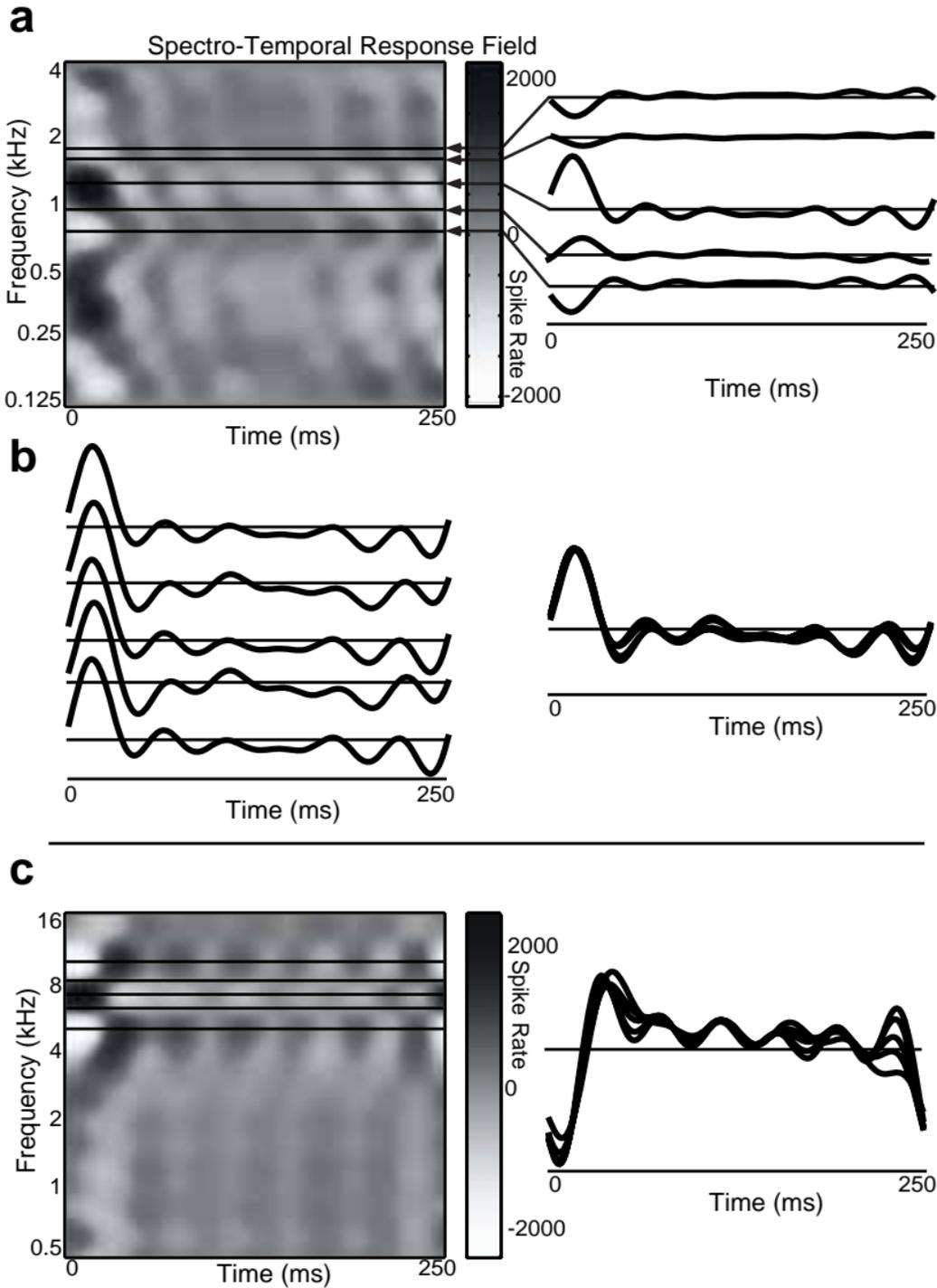

Figure 7

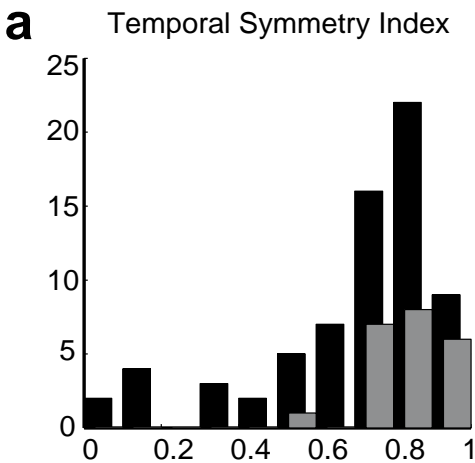

**a** Temporal Symmetry Index

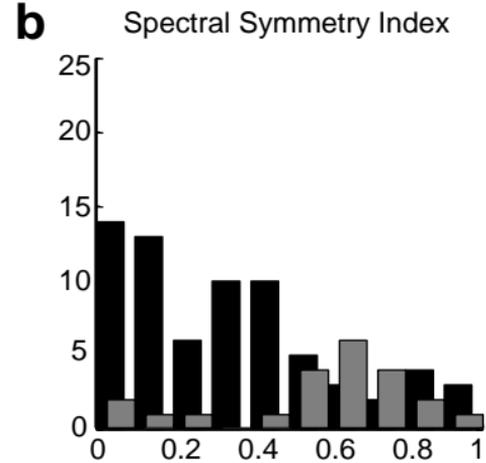

**b** Spectral Symmetry Index

Figure 8

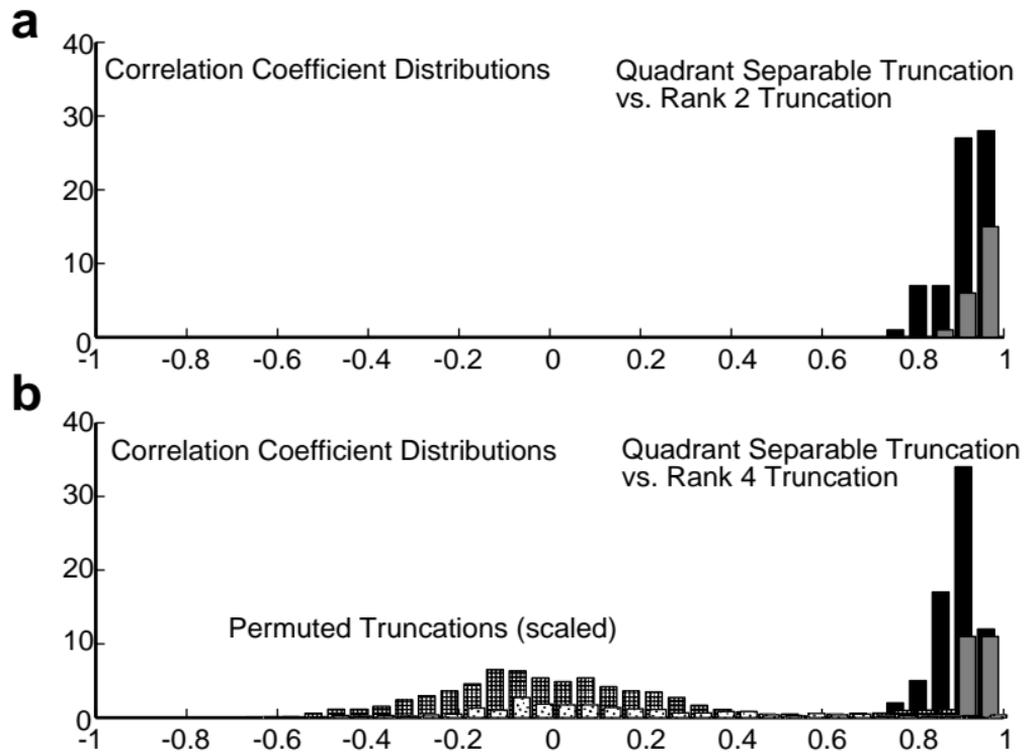

Figure 9

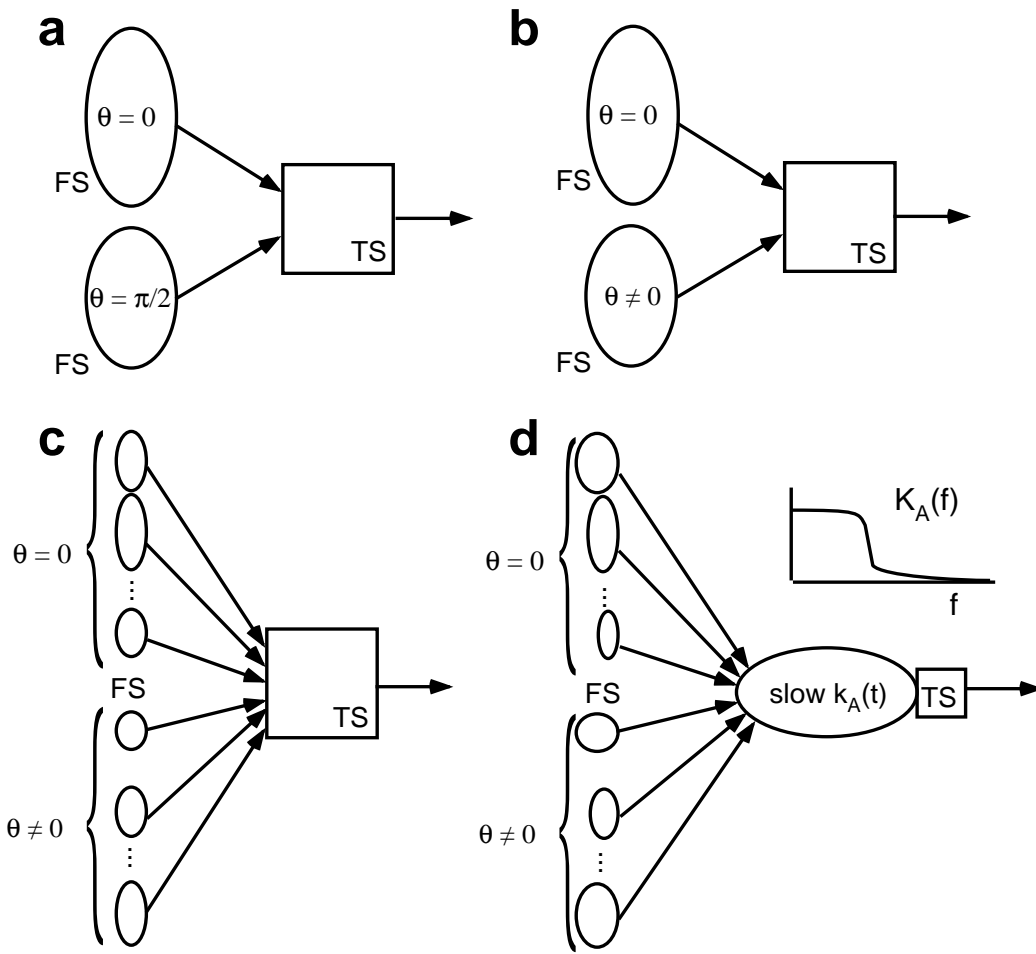

Figure 10

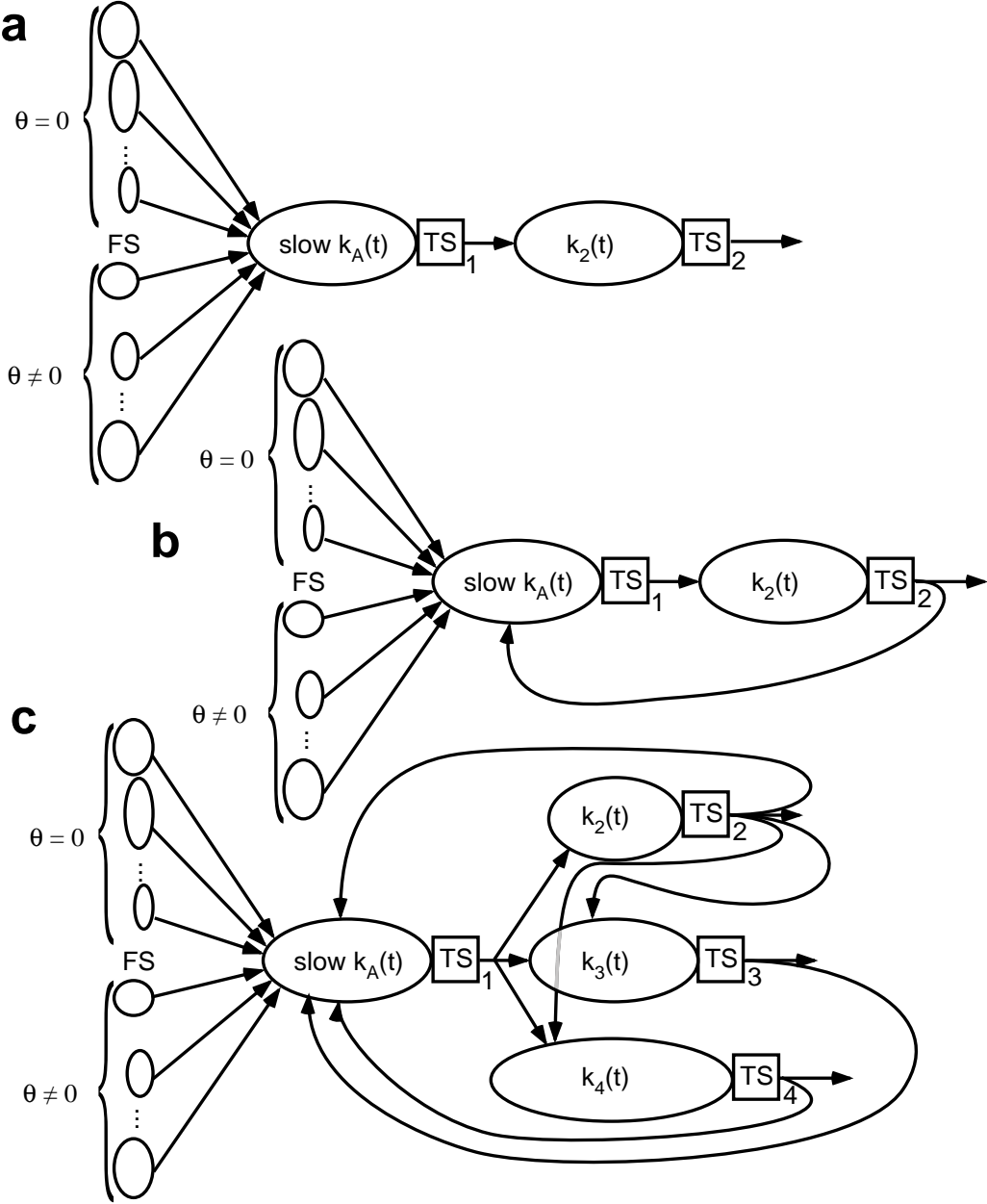

Figure 11